\documentclass[a4paper,fleqn,usenatbib]{mnras}

\usepackage{newtxtext,newtxmath}
\usepackage{mathptmx}
\usepackage{txfonts}
\usepackage{lmodern}
\usepackage[T1]{fontenc}
\usepackage{ae,aecompl}
\usepackage{rotating}
\usepackage{footnote}
\usepackage{times}
\usepackage{graphicx}	% Including figure files
\usepackage{amsmath}	% Advanced maths commands
\usepackage{amssymb}	% Extra maths symbols

\def\mnras{MNRAS}
\def\apj{ApJ}
\def\apjs{ApJS}
\def\aap{A\&A}
\def\aaps{A\&AS}
\def\aj{AJ}

\def \st{\ifmmode{^{\mathrm{st}}}\else{${^{\mathrm{st}}}$}\fi}
\def \nd{\ifmmode{^{\mathrm{nd}}}\else{${^{\mathrm{nd}}}$}\fi}
\def \rd{\ifmmode{^{\mathrm{rd}}}\else{${^{\mathrm{rd}}}$}\fi}
\def \th{\ifmmode{^{\mathrm{th}}}\else{${^{\mathrm{th}}}$}\fi}
\newcommand{\vel}{\rm km s$^{-1}$}

\newcommand{\NII}{[N~{\sc ii}]}

\title[High velocity knots in the outburst of the PN Hb4]
{High velocity string of knots in the outburst of the Planetary Nebula Hb4}

\author[Derlopa et al.]{S. Derlopa,$^{1,2}$\thanks{E-mail: sophia.derlopa$@$noa.gr}
S. Akras,$^{3}$ P. Boumis,$^{1}$\thanks{E-mail: ptb$@$astro.noa.gr} W. Steffen,$^{4}$\\
$^{1}$Institute for Astronomy, Astrophysics, Space Applications
and Remote Sensing, National Observatory of Athens,
15236 Penteli, Athens, Greece\\
$^{2}$Department of Physics, University of Athens, Athens, Greece\\
$^{3}$Observat\'orio Nacional/MCTIC, Rua Gen. Jos\'{e} Cristino, 77, 20921-400, Rio de Janeiro, Brazil\\
$^{4}$Instituto de Astronom\'ia, Universidad Nacional Aut\'onoma de M\'exico, Ensenada 22800, Baja California, Mexico\\}

\date{Accepted XXX. Received YYY; in original form ZZZ}

\pubyear{2018}

\hypersetup{draft}
\begin{document}
\label{firstpage}
\pagerange{\pageref{firstpage}--\pageref{lastpage}}
\maketitle

\begin{abstract}
 The bipolar collimated outflows of the Hb4 Planetary Nebula (PN) exhibit an evident decrease in their expansion velocity with respect to the distance from the central star. So far, similar velocity law has also been found in Herbig-Haro objects. The interpretation of this peculiar velocity law and the classification of the outflows is the main focal point of this paper. High dispersion long-slit echelle spectra along with high resolution images from Hubble Space Telescope ({\it HST}) are applied in the astronomical code SHAPE in order to reproduce a three-dimensional morpho-kinematicalal model for the core and the bipolar outflows. Its central part shows a number of low-ionization filamentary structures (knots and jets) indicative of common-envelope PNe evolution and it is reconstructed assuming a toroidal structure. The high-resolution {\it HST} {\NII} image of Hb4 unveils the fragmented structure of outflows. The northern and southern outflows are composed by four and three knots, respectively, and each knot moves outwards with its own expansion velocity. They are reconstructed as string of knots rather than jets.This string of knots is formed by ejection events repeated every 200-250 years. Hb4 displays several indirect evidence for a binary central system with a [WR] companion evolved through the common envelopes channel.The observed deceleration of the knots is likely associated with shock collisions between the knots and the interstellar medium or nebular material. 

%This is a simple template for authors to write new MNRAS papers.
%The abstract should briefly describe the aims, methods, and main results of the paper.
%It should be a single paragraph not more than 250 words (200 words for Letters).
%No references should appear in the abstract.
\end{abstract}


%%%%%%%%%%%%%%%%%%%%%%%%%%%%%%%%%%%%%%%%%%%%%%%%%%

%%%%%%%%%%%%%%%%% BODY OF PAPER %%%%%%%%%%%%%%%%%%

\section{Introduction}
\label{intro}

In a general description, a Planetary Nebula (PN) is an emission nebula consisted of an expanding bright shell of gas ionized by the strong UV radiation field of the central white dwarf. It is formed when an evolved, low-to-intermediate mass star (1-8 M$_{\sun}$) expels its outer layers in the Asymptotic Giant Branch (AGB) phase, in the form of a slow ($10~$\,km\,s$^{-1}$) and dense stellar wind (10$^{-4}$ M$_{\sun}$) at a high mass loss rate. Then, this material interacts with a fast ($\sim$1000 \vel) and tenuous stellar wind resulting in the formation of spherical PNe based on the assumption that both are spherically symmetric (Interacting Stellar Wind Model - ISWM; \citealt{KWO1978}).

However, the majority of PNe exhibit a diversity of aspherical structures (e.g. elliptical, bipolar; \citealt{SCH1992}; \citealt{BOUMIS2003, BOUMIS2006}; \citealt{SAH2011};\citealt{WEI2016}). The interaction between an AGB wind with high density difference between the poles and the equator with a spherically symmetric fast wind can explain the formation of aspherical PNe (Generalized ISWM; \citealt{BAL1987}; \citealt{ICK1988}). Common envelopes around binary systems provide the necessary density difference between the polar and equatorial directions on the AGB wind (e.g. \citealt{LIV1988}) ensuing the formation of bipolar PNe (e.g. \citealt{FRA2018}; \citealt{GARSEG2018}).

\par
An interesting case of PN due to its peculiar morphology and kinematical properties is Hb4 (G003.1+02.9; $\alpha_{2000}$: $17^{\rm h} 41^{\rm m} 52.7^{\rm s}$, $\delta_{2000}$: $-24\degr 42' 08.0''$). It is located in the Galactic Disk, with an estimated distance of 2.88~$\pm0.86$ kpc \citep{FRE2016}. 
{\it Gaia} DR2 gives a parallax of 0.435$\pm$0.189 mas \footnote{the {\it  astrometric excess noise} is lower than 1 which implies a reliable parallax measure, but due to the high fractional parallax error, the inverse parallax does not provide a reliable distance estimate.} (\citealt{GAI2018}). \citet{BAILJON2018}, using a statistical approach, estimate its geometric distance to 
2.55 kpc with 1$\sigma$ minimum and maximum bounds of 1.57 and 5.29 kpc, respectively. 

\citet{SAH2011} classify Hb4 as a multi-polar nebula with irregular structure and barrel inner region. At the outer parts, it also shows a pair of collimated, detached jets or elongated knots moving with a velocity of $\sim$150 \vel (\citealt{LOP1997}). \citet{RAB2003} reported a faint secondary bipolar structure, close to the central part, aligned with the minor axis, a feature that in the end implies poly-polarity for this nebula (see their, figure 1).

Hb4 is also identified as Type I PN in Peimbert's scheme due to its high {\bf N/O} abundance ratio \citep{PEN2017}. The abundance discrepancy factor of singly ionized oxygen is found to be equal to 3.7 (\citealt{GARROJ2013}), which may be indicative of a binary central star (\citealt{COR2015}). Its nucleus is classified as hydrogen-deficient star of [W03] class by \citet{ACKNEI2003} and of [WC4] class by \citet{GOR2004}. A possible link between jets and knots with binary systems or [WR] central stars in PNe has also been proposed (\citealt{MIS2009}). 
 
\par

The most striking characteristic of Hb4 on which this paper focuses, is the pair of bipolar collimated outflows (\citealt{LOP1997, HAJ1997}), protruded from both sides of the main plane of the nebula, displaying a spectrum of low-ionization structures (LISs; \citealt{GON2001}, \citealt{AKRGON2016}) relative to the rest of the nebula \citep{COR1996}. These two outflows exhibit  H$\alpha$/\NII~6584 $\AA$ line ratio of $\sim$1.10 (northern) and $\sim$1.20 (southern), in contrast with the value of 3.29 of the core \citep{LOP1997}, which is expected, since this ratio is much stronger in the outflows than in the core, as already has been mentioned in low- ionization structures \citep{AKRGON2017}. According to \citet{LOP1997}, the systemic heliocentric radial velocity is $V_{\rm sys}$ =$-$58.9~\,km\,s$^{-1}$, while the expanding velocity of the shell that surrounds the core of Hb4 is $V_{\rm exp} \simeq 21.5~$\,km\,s$^{-1}$. This is in good agreement with the calculated values of $V_{\rm exp}$ = 23$\pm 2~$\,km\,s$^{-1}$ \citep{ROB1982} and $V_{\rm exp}$~=~23$\pm 4~$\,km\,s$^{-1}$ \citep{DAN2014}, respectively.
\par
\citet{LOP1997}, in their kinematical analysis on Hb4 modelled the outflows as bow-shocks. 
On the other hand, \cite{DAN2014} defined these structure as "point-symmetric thin knots". \citet{HAJ1997} suggested that the southern outflow of Hb4 could be defined as jet, while for the northern structure they proposed the term FLIERS (fast, low-ionization emission regions; \citealt{BAL1993}) due to the knots' size, peculiar velocities and almost symmetric placement on opposite sides of the central star.
\par
FLIERS are low ionization microscopic structures ($\sim$1\arcsec) within PNe in the shape of knots,  jets, filaments, etc. They are usually found in pairs, exhibit equal but opposite velocities with respect to the nebula's systemic velocity. According to \cite{GON2001}, FLIERS fall into the general category of LISs, i.e. small scale, low-ionization structures which cover a wide range of expansion velocities from few tens to hundreds of \,km\,s$^{-1}$. Therefore, considering their kinematical behaviour, LISs can be separated into FLIERS as we mentioned above, BRETS (bipolar rotation episodic jets; \citealt{LOP1995}), or SLOWERs (slow moving, low ionization emitting regions; \citealt{PER2000}). The nature of LISs is still an open question. They exhibit very similar morphologies, kinematicals and emission line spectra to many Herbig-Haro (HH) objects, which are primarily heated and excited by collisions \citep{BAL1993, BAL1998}. It has been recently shown that a combination of photo-ionization by the strong UV radiation of the central stars and shock interactions of the LISs with the surrounding nebular material can explain their spectral characteristics (e.g. \citealt{RAG2008}; \citealt{AKRGON2016}; \citealt{AKRGON2017}).

\par
In this paper new spectroscopic data of Hb4 along with high resolution {\it HST} optical images are applied in the astronomical code SHAPE with the aim to clarify the morphology and kinematical behaviour of the two bipolar outbursts in question, through the produced 3D model of the core and the outflows of Hb4. Our focus on the outflows is motivated by the observational spectrum in which an unusual decrease of the expansion velocity is observed as a function of distance from the core. This analysis can give feedback for the origin of these outflows, and subsequently for the mass loss mechanism during the evolution of the nebula that produced them. The observations and data analysis are described in Section 2. SHAPE modeling and the results of the morpho-kinematical models are presented in Section 3 and 4 respectively. The results of this work are discussed in Section 5, and we end up with the conclusions in Section 6. 
 
\section{Observations}

\subsection{High resolution imaging}

The optical image used (Figure \ref{fig:HST4SLITSWfigure}) was obtained from the Mikulski Archive for Space Telescopes, from observations made with the Hubble Space Telescope ({\it HST}) on October 28\th\ 1996 \citep{BOR1996}. The detector was the Wide Field Planetary Camera 2 (WFPC2) with 800$\times$800 pixels, each one with 15 $\mu$m size. The field of view was 2\arcmin.5 $\times$ 2\arcmin.5 and the image scale 0\arcsec.1 pixel$^{-1}$. For the needs of the presented probe we used the narrow-band filter F658N ($\lambda {\rm c}$=6591~$\AA$) with an exposure time of 400\,sec.

\subsection{High-dispersion long slit spectroscopy}

The observed long slit echelle spectra were obtained in March 2015 at the 2.1 m telescope in San Pedro Martir National Observatory, Mexico, with the Manchester Echelle Spectrometer (MES-SPM; ~\citealt{Meaburn2003}). This spectrometer has no cross dispersion. H$\alpha$ 6563 $\AA$ and {\NII} 6548,6584 $\AA$ nebular emission lines were isolated by the 87th echelle order. Since the H$\alpha$ emission line features have similar morphology to that of the {\NII}, we present here only the latter for comparison reasons with the data of \citet{LOP1997}.
\par 
The Marconi E2V 42-40 camera, with 2048 $\times$ 2048 squared pixels, each with 13 $\mu$m was the detector. Two times binning was used in both the spatial and spectral dimensions which resulted in 0\arcsec.35 pixel$^{-1}$ spatial scale. The 1076 increments gave a total projected slit length of 5\arcmin.12 on the sky. The slit width used was 150 $\mu$m (10 \,km\,s$^{-1}$ and 1\arcsec.9 wide), and was oriented at Position Angle (hereafter P.A.) of 22$\degr$ along the axis that connects two outflows, as seen in Figure \ref{fig:HST4SLITSWfigure}(a) (slit S4). The exposure time for each spectrum in each filter was 1800\,sec.
The data were analysed in the typical way using the IRAF software package. The spectra were calibrated in heliocentric radial velocity ($V_{hel}$) to $\pm2.6$ \,km\,s$^{-1}$ accuracy against spectra of a thorium/argon lamp. The atmospheric seeing was varying from 1 to 1.5 arcsec during the observations.

 \section{SHAPE Modelling}
 \label{SHAPE}

 Most morphological information of PNe comes from imaging observations, but can be complemented with kinematicals that is obtained from the spectrum through the produced Position-Velocity (PV) diagrams.
 For the three-dimentional visualization of Hb4, we used the code SHAPE (\citealt{STE2006}; \citealt{STE2011}) which has been extensively used to study the morpho-kinematical structure of planetary nebulae (e.g. \citealt{AKRSTE2012, CLY2015}; \citealt{AKRCLY2016}). SHAPE provides a list of tools (structural or physical) that allow us to construct the 3D morphology and reproduce the kinematical structure of an extended object, in this case a PN. The synthetic output image and PV diagram generated by our model are qualitatively compared with the observational data. The modification of the model structure continues until the synthetic image and PV diagram reproduce adequately the observed images and PV diagrams respectively.

\section{Results}

\begin{figure*}
\begin{center}
 \includegraphics[scale=2]{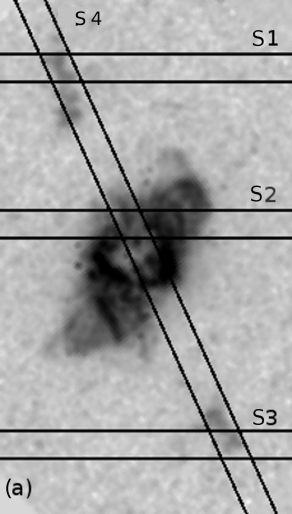}
 \includegraphics[scale=0.238]{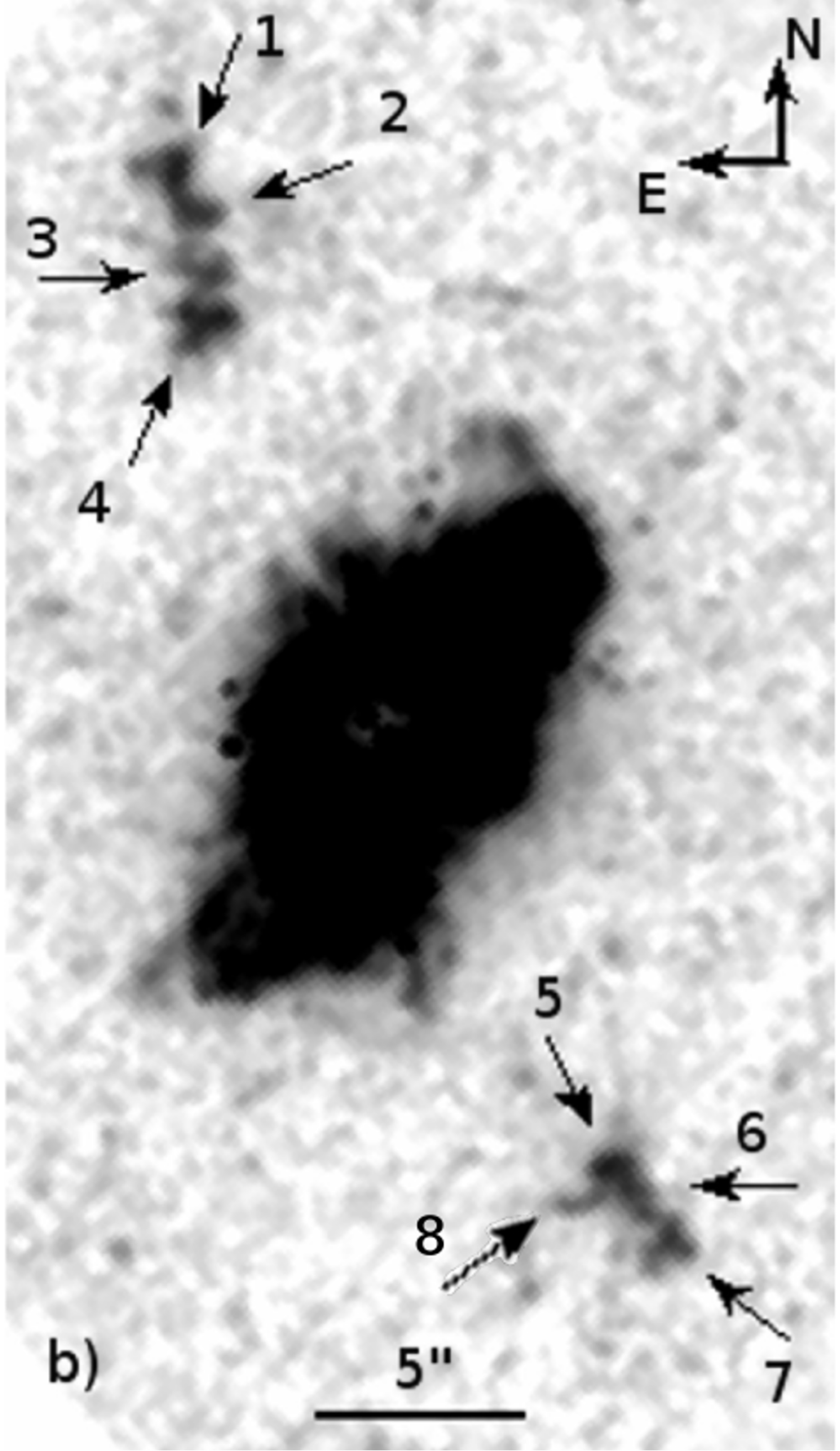}
 \caption{{\it HST} image of Hb4 in \NII~6584 emission line in low and high greyscale representation (logarithmic scale). a) Emphasis is given on the central main ring and the elliptical structure around it. Also displayed the three parallel slits of \citet{LOP1997} (S1-S3) along with the slit position of our observations (S4). b) Labelled knots on both sides of the central PN included within the slit S4, apart from knot 8 (see Section \ref{Discussion}).}
    \label{fig:HST4SLITSWfigure}
    \label{fig:knotsAfigure}
\end{center}
 \end{figure*}

The {\it HST} \NII~6584 image in Figure \ref{fig:HST4SLITSWfigure} displays the morphology of Hb4. Figure \ref{fig:HST4SLITSWfigure}(a) clearly shows the central bright ring of the nebula, the fainter roughly elliptical structure that surrounds that annular core, as well as a number of filaments and low-ionization structures. The three parallel slits positions of \citet{LOP1997} (S1-S3) oriented in the east-west direction and that of our observations at a P.A. of 22$\degr$ (S4) are also displayed. Figure \ref{fig:knotsAfigure}(b) emphasizes on the collimated outflows which, in Section \ref{knots}, are characterized as "knots".
\par
The elliptical structure, inside of which the main core is located, consists of a major and a minor axis, with projected lengths of approximately 15\arcsec and 6\arcsec, respectively. This structure in the periphery of the main ring, has a non-uniform shape which may indicates (a) strong stellar winds interaction with the interstellar/circumstellar medium or (b) an explosion preceded the final ejection of the gaseous shell.
\par
The bipolar outflows, which is the main subject of this paper, are ejected on both sides of the nebula, and are not perfectly aligned with respect to each other. The direction away from the central star of the northern outflows is tilted by approximately $\sim$5$\degr$ with respect to that of the southern one, in agreement with \citet{LOP1997}.
\par 
For the classification of the fast moving outflows two cases are considered: a) as jets and b) as string of knots. In both assumptions, we consider the same model for the core of the nebula, which we present directly in the following Section.

\subsection{Modelling the Hb4 core}
 \label{Hb4 core}
 
In order to represent the outflows of Hb4 by the use of the code SHAPE, it was necessary to also reconstruct the main core of the nebula, in order the core to constitute the point reference for the distances of the knots from the central star.
\par
The best-fitted model deduced from SHAPE, reproduces the core of Hb4 as a toroidal component. 
Figure \ref{fig:1c}(a) displays the mesh structure of the synthetic torus positioned on the {\it HST} image along with the observational slit position (two vertical lines of 1\arcsec.9 width). The image scale is the same as that in Figure \ref{fig:HST4SLITSWfigure}(b). There are also displayed the two groups of knots, which are described in Section~\ref{knots}, along with the symmetry axis perpendicular to the plane of the core (orange line). The slit is at a PA of 22$\degr$ with respect to the North, while -as derived from our SHAPE model- the major axis of the torus is at an angle of 107$\degr$ relative to the North, the angular size of the torus is 7 arcsec and its mean expansion velocity is $14\pm5$\,km\,s$^{-1}$.
\par
Figure \ref{fig:PV_core1afigure}(a) displays the observational PV diagram of Hb4 in the \NII~6584 emission line, in a low contrast that features the core's spectrum. Figure \ref{fig:PV_corecol1afigure}(b) illustrates the synthetic PV diagram of the torus obtained with SHAPE, which reproduces very well the observational PV. The colored areas in the synthetic spectrum represent the northern and southern parts of the core which are red-shifted and blue shifted, respectively. 

Having the representation of the nebula's core, we continue with the two scenarios for the clarification of the outflows.

\subsection{Outflows as "Jets"}
\label{Hb4 jets}
\par
What makes Hb4 intriguing is that, according to the observational PV diagram (Figure \ref{fig:pvfigure}(b), grey-scale representation), the most distant parts of the outflows seem to move outward more slowly than the closer ones with respect to the main core of the PN, which does not agree with previous finding in PNe (e.g. \citealt{RIE2002}; \citealt{VAY2009}) or either from jets (e.g. \citealt{COR1999}; \citealt{GARSEG1999}). 

Nevertheless, a similar "abnormal" behaviour is reported in \cite{DEV1997} where the observed outflows are eventually described as a chain of Herbig-Haro objects of complex morphologies, and radial velocities which systematically decline with increasing distance from the core of HH34 IRS. A similar expansion law -- exponentially decrease with the distance from the central star -- has reported in IRAS 18113-2503 (\citealt{ORO2018})  and it is attributed to the interaction between the outflow and the circumstellar envelope decelerating the gas.

Therefore, in the case of Hb4, the first scenario of uniform jets with a linear increase of velocity with the distance from the main core fails to reproduce the observations. This  result reinforces the second scenario that the outflows are likely separated substructures, each one of which has jumped out of the core with its own velocity at different explosion events.

\subsection{Outflows as "knots"}
\label{knots}

The second hypothesis for the morphology of the outflows, is to consider them as string of separated knots.
\par
The first indication for this, was the difference between the substructures that were captured in our slit position from the observations of 2015, and those that were present in the slits positions in the paper of \citet{LOP1997} (see Figure \ref{fig:HST4SLITSWfigure}(a)). We noticed that the microstructures presented in our observations, were absent from \citet{LOP1997}. This suggests that the outflows are likely not uniform structures, but depending on the slit position and orientation, different fragments shed from the PN are able to be detected.
\par
The second indication, which was quite unambiguous, came from the high resolution {\it HST} optical image. In Figure \ref{fig:knotsAfigure}(b), where Hb4 is illustrated in the \NII~6584 emission line, the fragmented nature of the outflows seems to be obvious. From that image -and the echelle spectra as explained in the following paragraph- the knots contained into slit S4 that were identified according to the available resolution, are knots 1-4 in the northeastern outflow and knots 5-7 in the southwestern outflow. Knot 8 which is also labelled on the image, is not included in slit S4, and it is explained analytically in Section \ref{Discussion}.
\par
The third indication came out from the contours in the observational PV diagrams. In Figure \ref{fig:contours1afigure}(a) the PV diagram of Hb4 is illustrated in the \NII~6584, along with the contour map. The contours parameters were chosen according to the counts of the knots, not the core's, which explains the difference in the contour curves we see in the plot. The formed contours on the outflows are an indirect but adequate proof that these outflows are not an undivided structure, but they consist of many individual substructures in different distances relative to the main star, and with different velocities. Again, the labelled vectors indicate the position of each knot -included into the slit S4- in the observational PV diagram.

\begin{figure}
\begin{center}
 \includegraphics[scale=0.223]{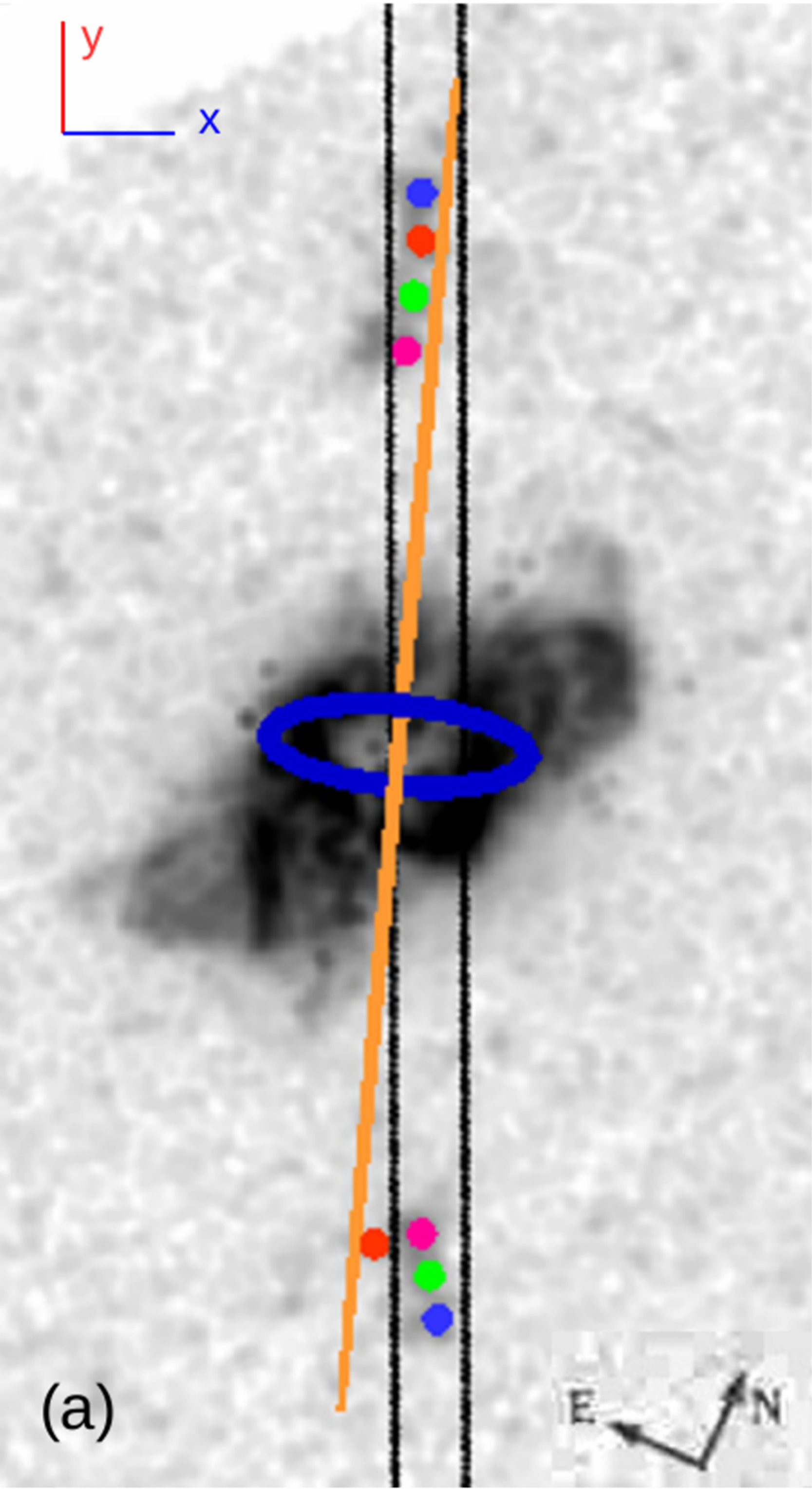}
 \includegraphics[scale=0.1415]{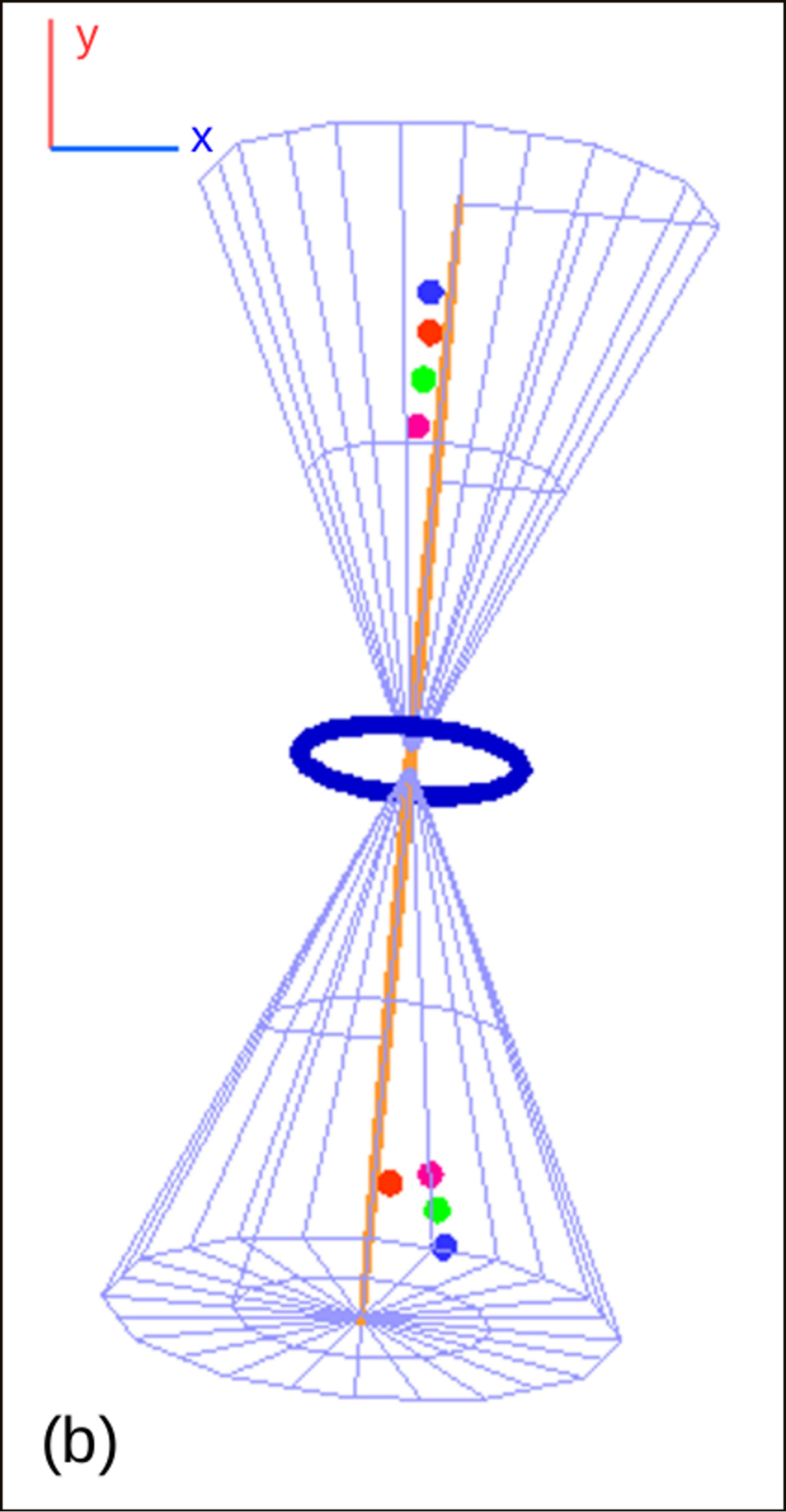}
 \caption{a) Superposition of {\it HST} \NII~6584 image along with the slit position and the produced SHAPE model of the core, the knots for Hb4 and the symmetry axis perpendicular to the plane of the core (orange line). The scale is the same as in Figure \ref{fig:knotsAfigure}(b). b) The same model without the {\it HST} image, and with the two cones which define the area of the knots' emission from the core.}
    \label{fig:1c}
    \label{fig:2d}
    
 \end{center}
  \end{figure}

   \begin{figure*}
  \begin{center}
   \includegraphics[scale=0.278]{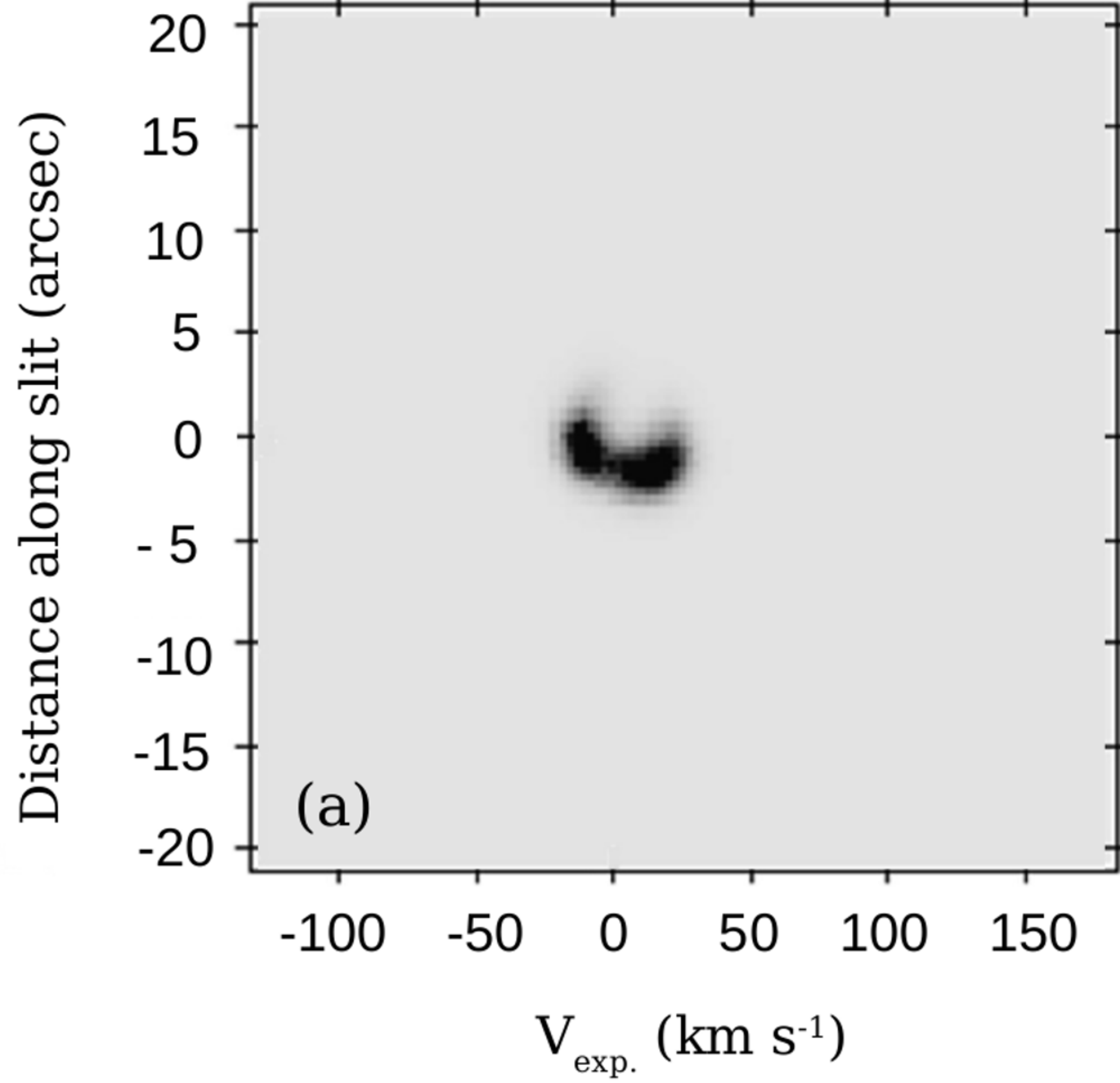}
   \includegraphics[scale=0.28]{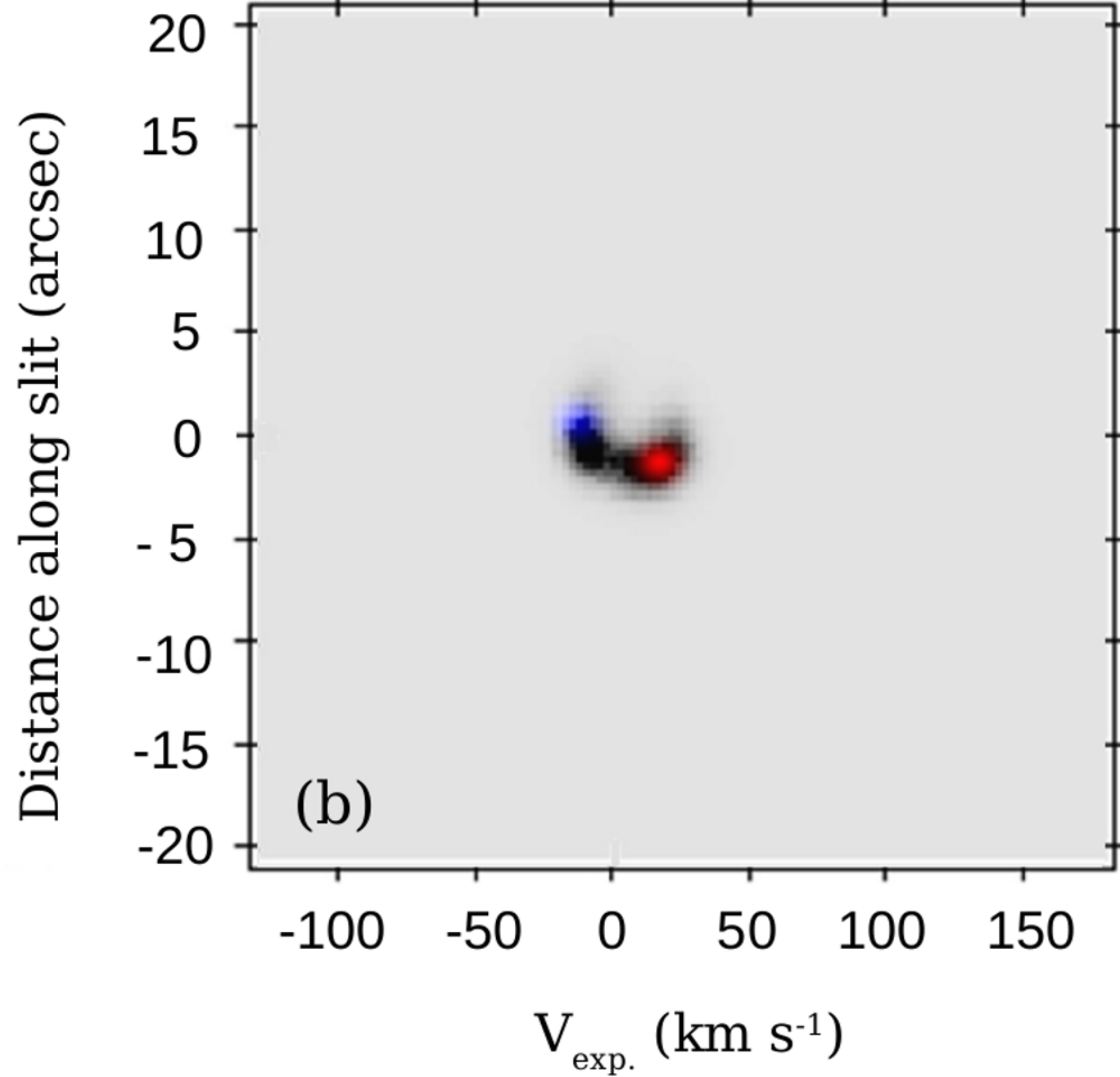}
   \caption{(a) Observational PV diagram of the core of Hb4 in the \NII~6584 emission line (slit S4). (b) The synthetic PV of the core obtained with SHAPE, overlaid on the observational one. The red and blue shifted regions of the core indicate the radial expansion of the core of Hb4.}
      \label{fig:PV_core1afigure}
      \label{fig:PV_corecol1afigure}
  \end{center}
   \end{figure*}
   
   \begin{figure*}
  \begin{center}
   \includegraphics[scale=0.30]{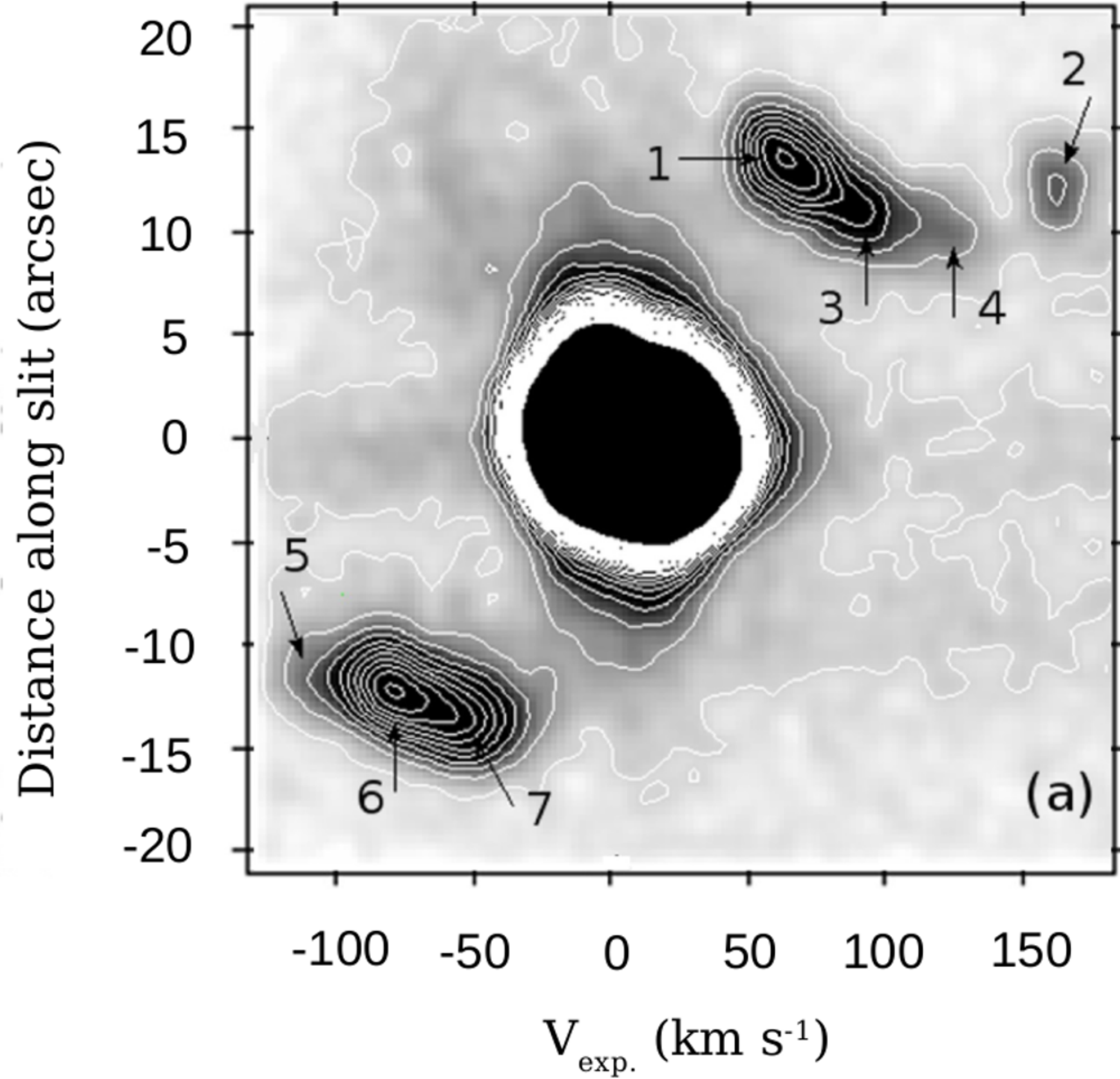}
   \includegraphics[scale=0.198]{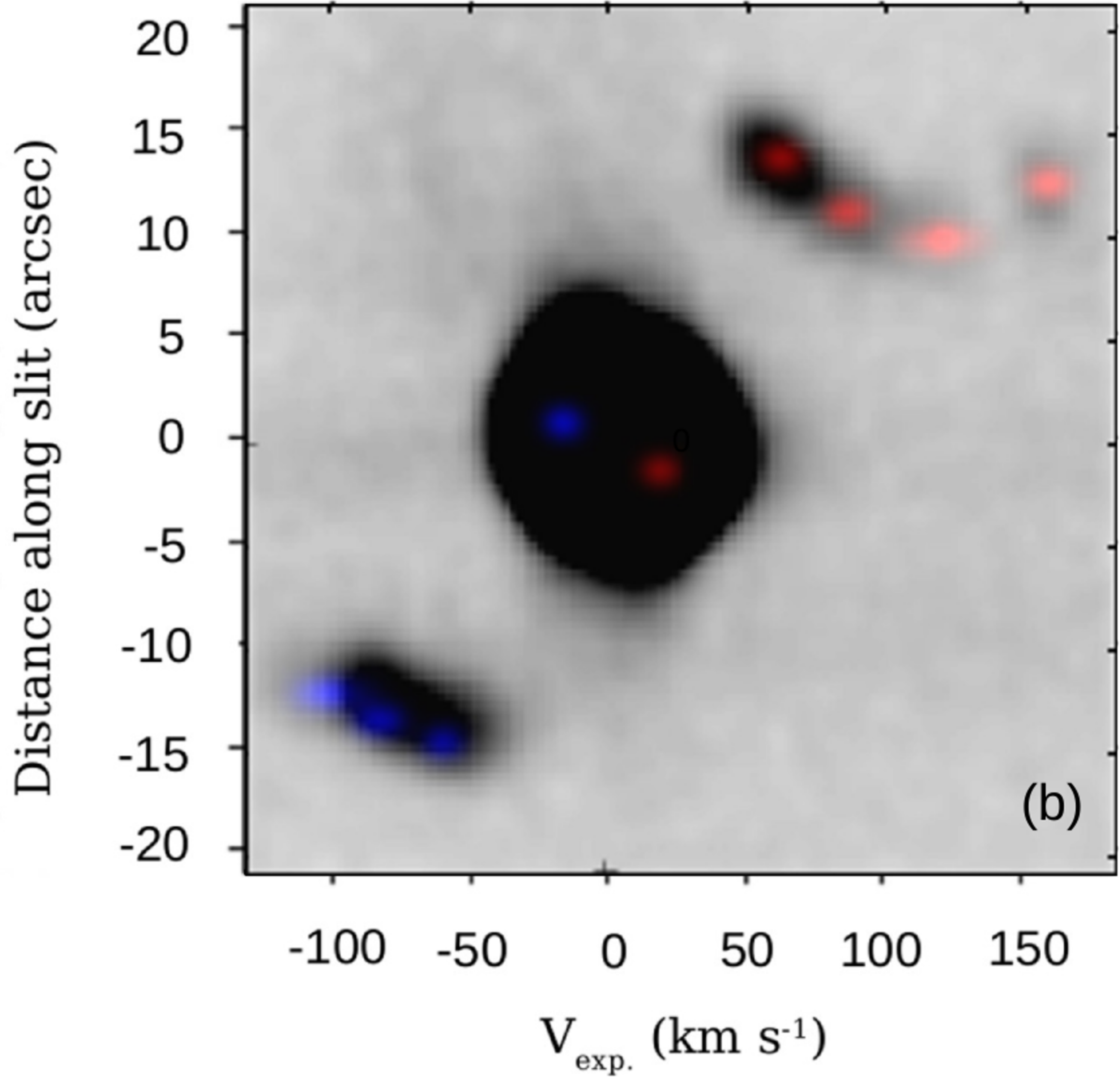}
   \caption{(a) Observational PV diagram in the \NII~6584 emission line of Hb4 along with the contours lines and the positions of the knots included into the slit S4 (in linear scale). (b) Superposition of the observational and the, produced with SHAPE, synthetic PV diagram.}
      \label{fig:contours1afigure}
      \label{fig:pvfigure}
  \end{center}
  \end{figure*} 
  
  \begin{figure*}
  \begin{center}
   \includegraphics[scale=0.1542]{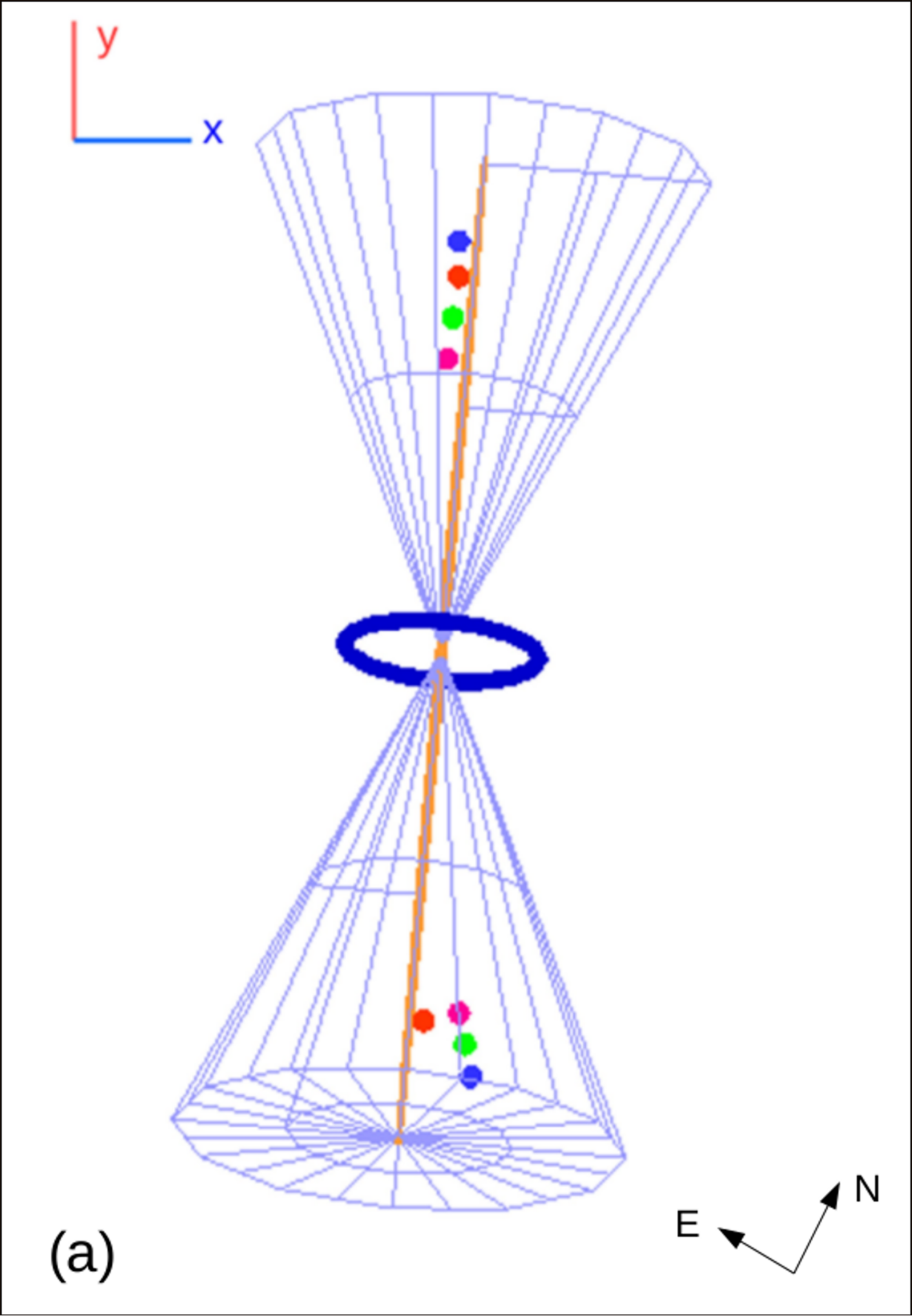}
   \includegraphics[scale=0.1502]{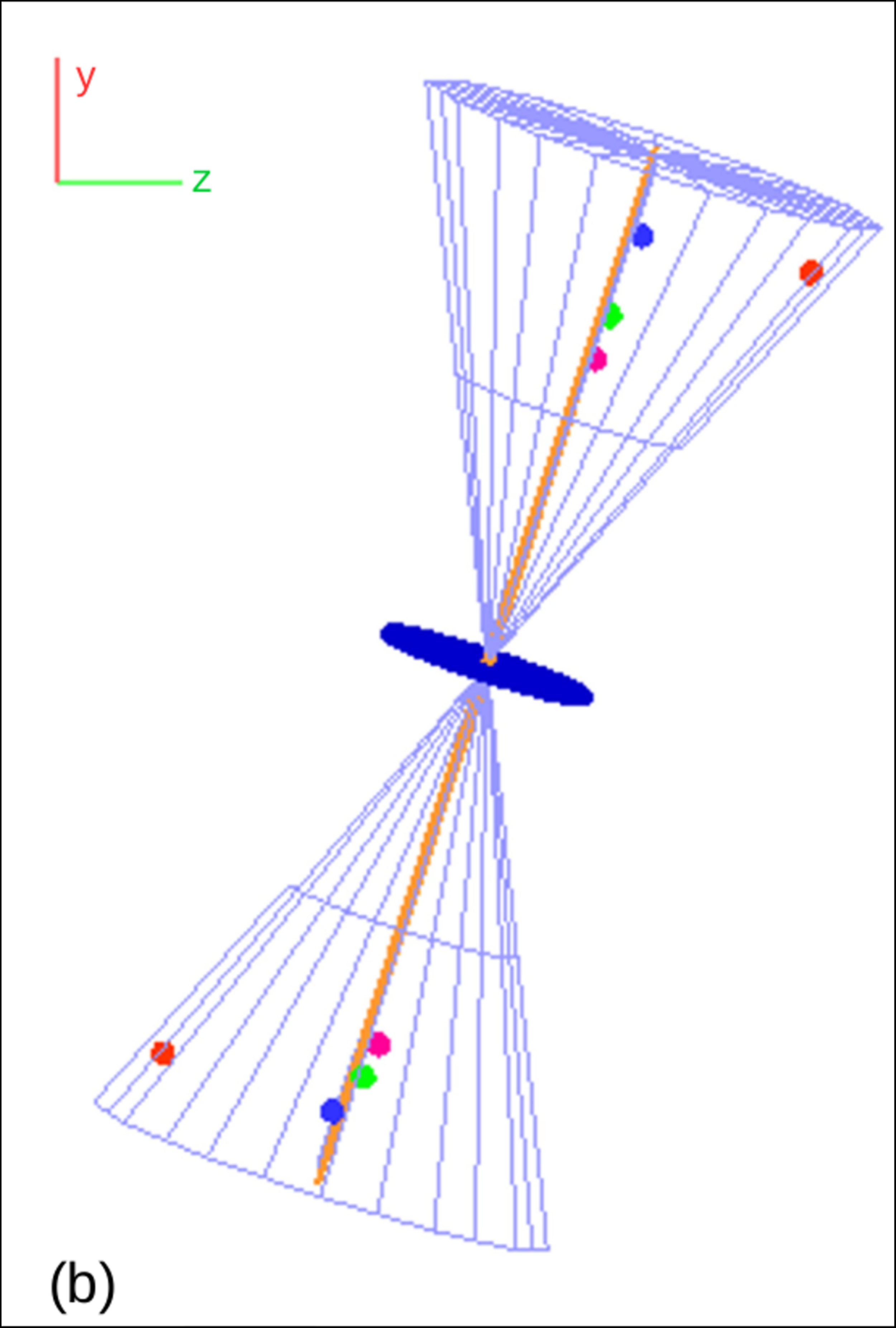}
   \includegraphics[scale=0.1515]{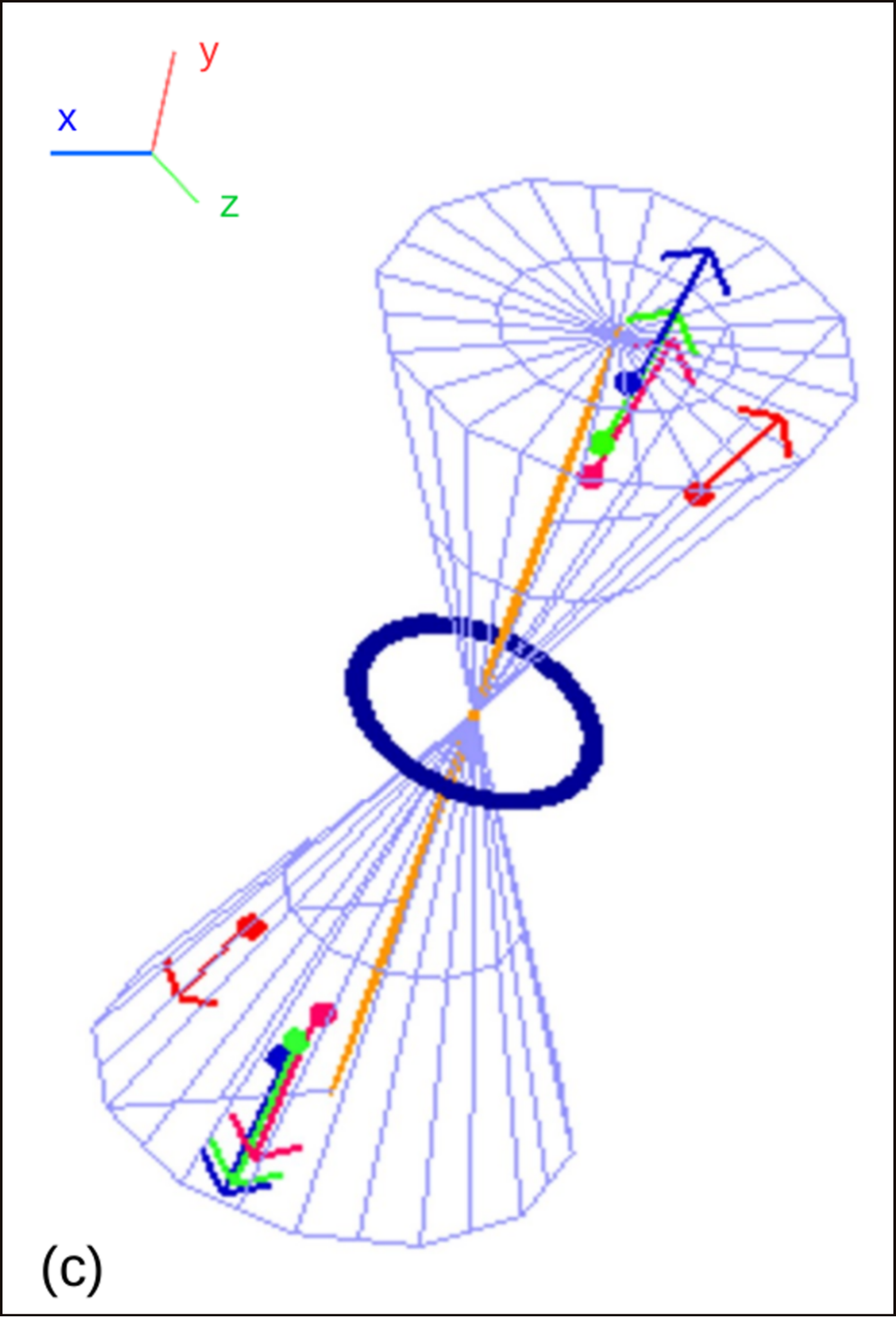}
   \includegraphics[scale=0.1537]{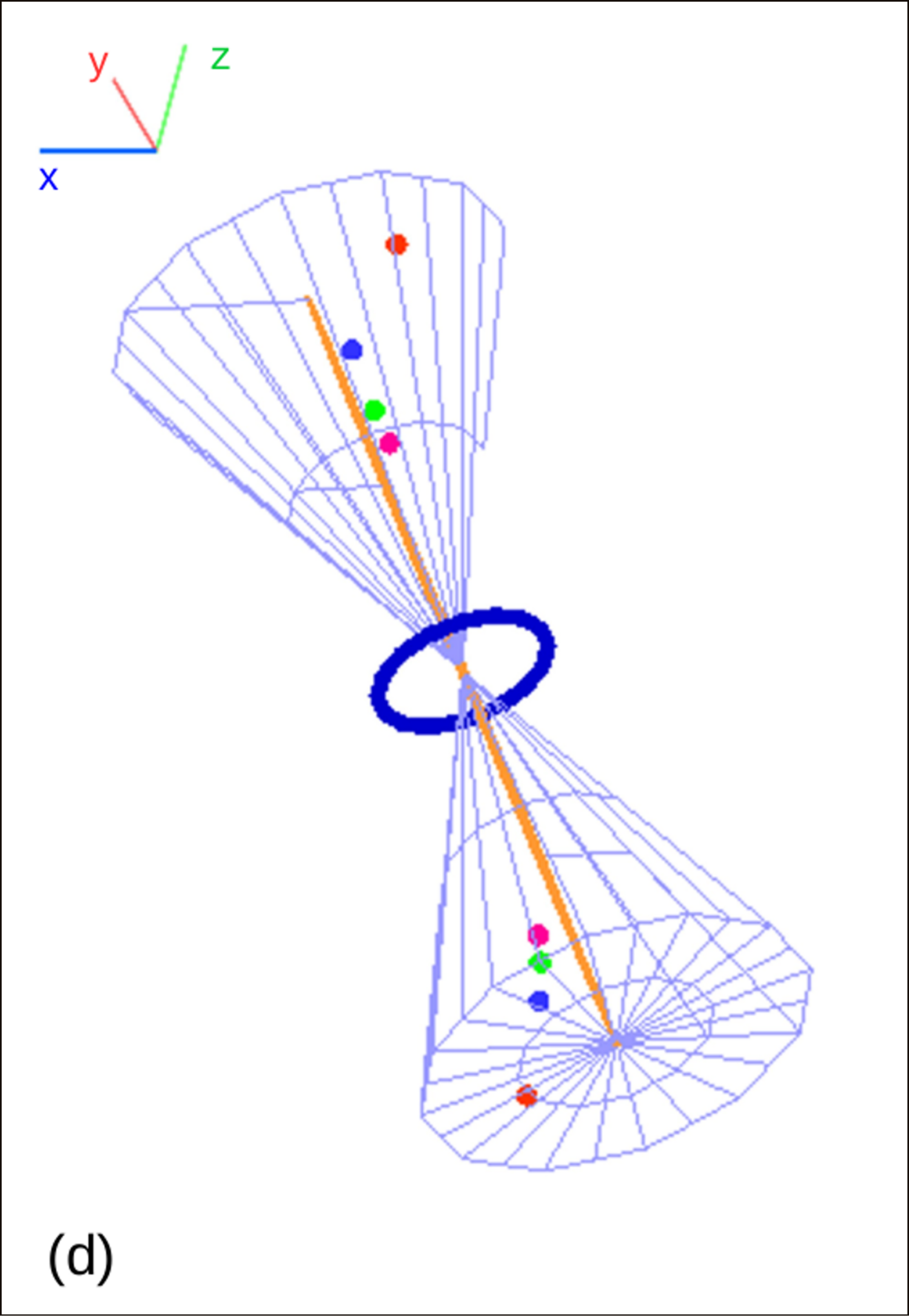}
   \caption{The produced SHAPE model of the core and the knots for Hb4, as it is displaced from four different angles. The orange line is the symmetry axis of the PN, perpendicular to the plane of the torus. The two cones demarcate the areas within which the knots are assumed to have been expelled from the core. In (c) image, the colored vectors represent the direction of Vout for each knot.}
      \label{fig:2g}
      \label{fig:3da}
      \label{fig:4c}
      \label{fig:5d}
   \end{center}
    \end{figure*}
    
 \begin{figure}
 \begin{center}
 \includegraphics[scale=0.15]{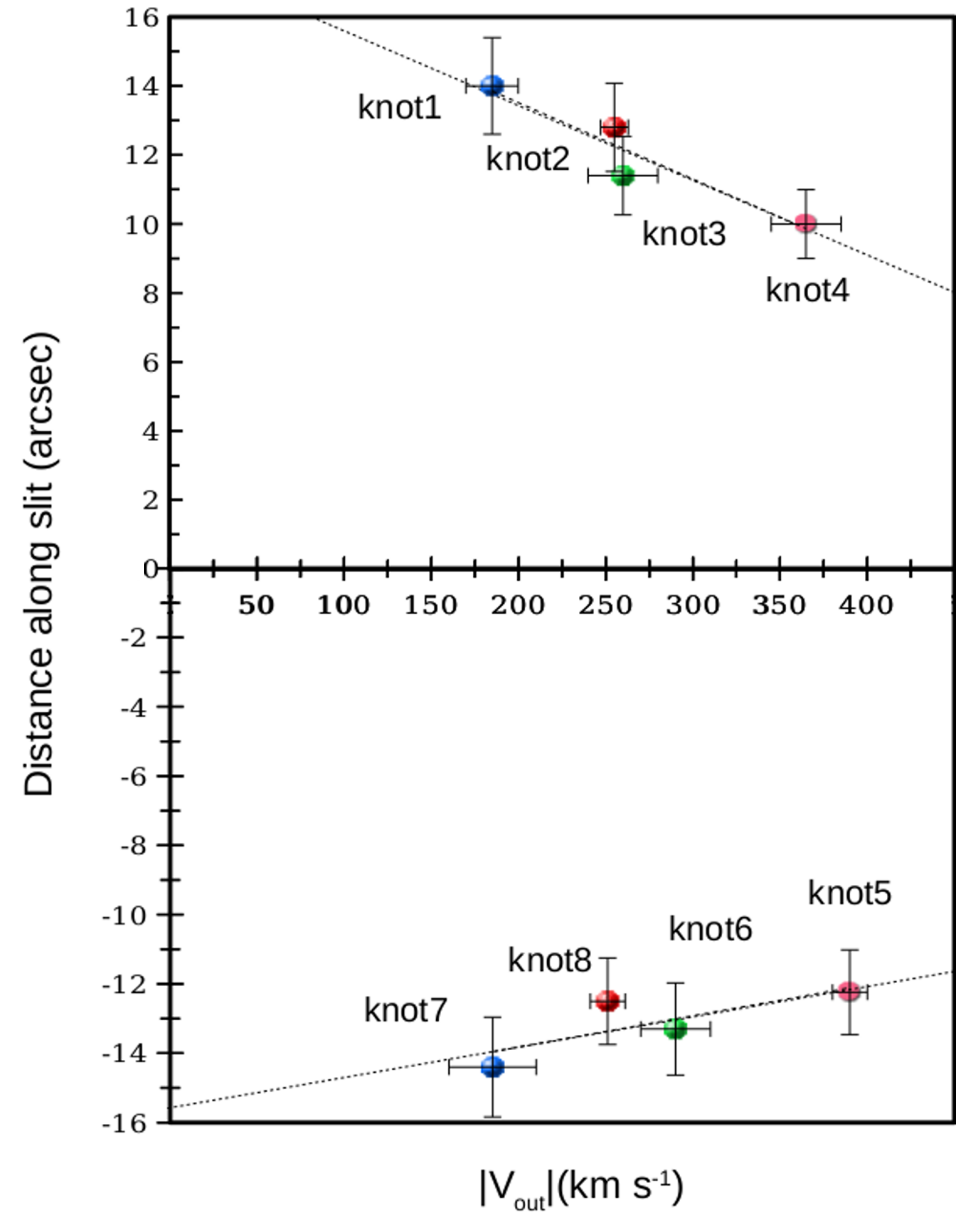}
 \caption{Plot of the  absolute outflow velocity of each knot with respect to its distance along the slit, relative to the core of Hb4. The colors of the knots are the same as in the SHAPE model. Knots with the same color constitute a pair of knots, considered to have been expelled at the same time from the core of Hb4. The dotted lines represent the best fitted lines obtained with the least-squares method.}
\label{fig:plot3_errors}
\end{center}
 \end{figure}

    \begin{figure}
      \begin{center}
      \includegraphics[scale=0.22]{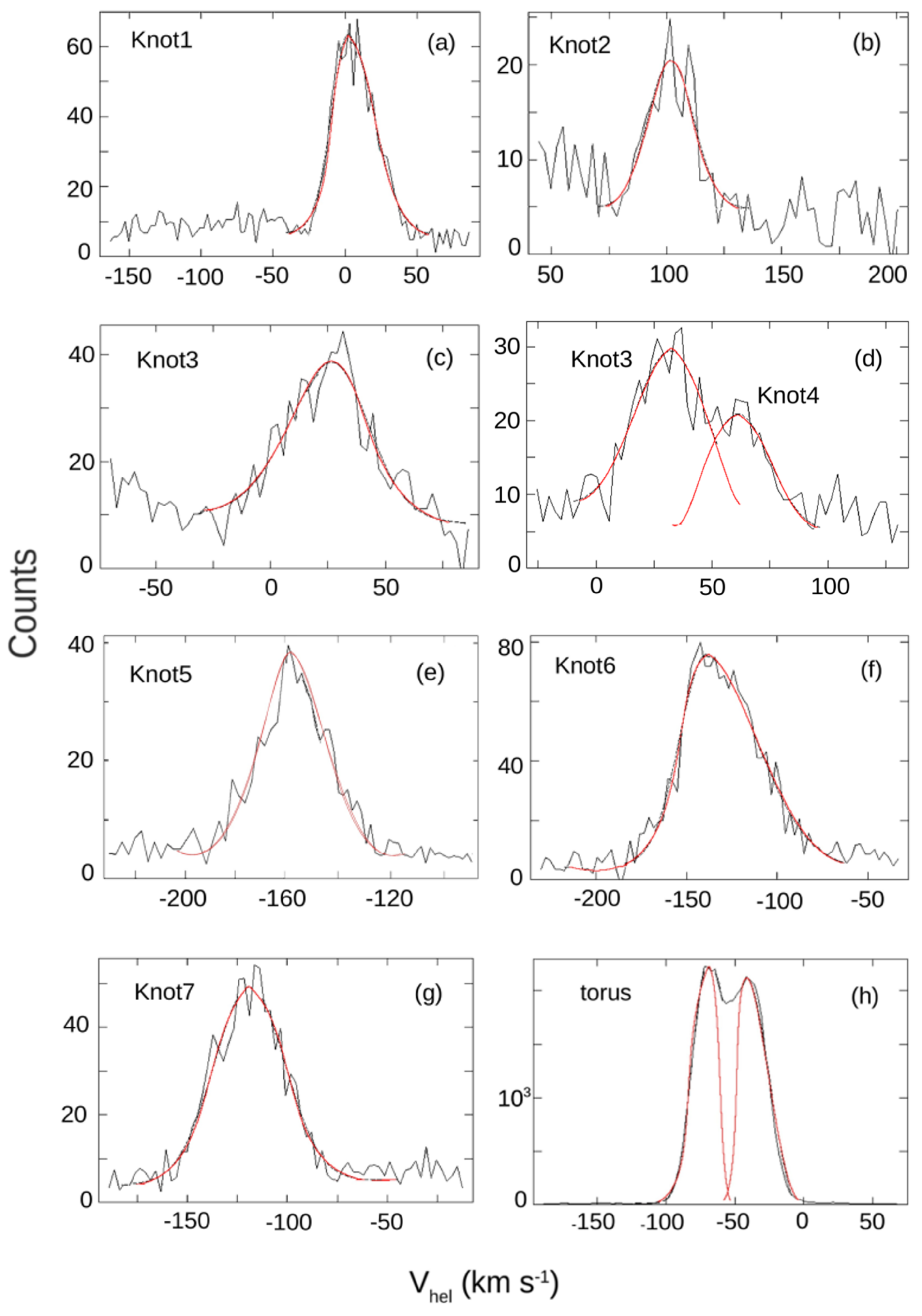}
      \caption{\NII~6584 emission line profiles of the knots identified in slit S4, along with the profile for the torus of the nebula. In (d) both knots 3 and 4 are presented due to the difficulty to separate them because they are very close to each other.}
     \label{fig:profiles_torus}
     \end{center}
      \end{figure}
  \par
Taking into account the above indications, the scenario of a string of knots is in favour. Therefore, having as a fact that the collimated outflows in Hb4 are fast moving detached knots receding on both sides of the core of the PN, we proceeded to the representation of the knots with SHAPE as small spheres, each one of which follows a linearly decreasing velocity law towards the distance from the central star.
\par
Figure \ref{fig:1c}(a) illustrates the superposition of the {\it HST} \NII~6584 image along with the 3D mesh of the final model of Hb4, where the outflows are considered as group of separated knots. Furthermore, it displays the axis of symmetry of the system perpendicular to the plane of the torus (orange line), as well as the model for the torus, as this was described in Section~\ref{Hb4 core}. All of these structures were produced with the SHAPE code.Information about the knots' position derived from the SHAPE model (distance from the nuclear center and projected angle of ejection with respect to the main axis) along with the velocities for the knots and torus, are presented in Table \ref{tab:Table1}. In this table, the expansion velocity $V_{\rm exp}$ = $V_{\rm hel}$ $-$ $V_{\rm sys}$, is derived from the observational data and is the velocity with which every structure of the PN is moving relative to its center, where we consider the velocity equal to zero. The orientation of $V_{\rm exp}$ is along the line of sight, i.e. inwards the page along the z axis (see Figure \ref{fig:1c}(a)). The velocity law in SHAPE is $V$ = $B(\frac{r}{ro})$, where r and ro are given in arcsec. In Table \ref{tab:Table1}, B stands for the outflow velocity ($V_{\rm out}$) given in km\,s$^{-1}$, is a velocity derived from the model and its direction is along the symmetry axis of the PN (see Figure \ref{fig:4c}(c)). The projection of $V_{\rm out}$, is the expansion velocity. In Figure \ref{fig:1c}(b) is displaced the same SHAPE model for the whole system along with the two cones, each one of radius $7$~\arcsec and height $16.5$ \arcsec. These cones indicate the region within which the emission of the knots is assumed to have taken place.
\par
For the better understanding of the knots' position in 3D space, four images are presented in Figure \ref{fig:2g}, where our model is viewed from four different angles. In Figure \ref{fig:2g}(a), Figure \ref{fig:1c}(b) is displaced again for comparison reasons, axis z points inwards the page. Figure \ref{fig:2g}(b) is like seeing Figure \ref{fig:2g}(a) from the right. Figures \ref{fig:2g}(c) and \ref{fig:2g}(d) present our model in free-form angles. The knots with the same color are assumed to consist pairs of emitted knots, as it is discussed in Section \ref{Discussion}. 

The synthetic PV which corresponds to this final model is shown in Figure \ref{fig:pvfigure}(b) in colored-scale, along with the observational PV in gray-scale. Indeed, this model gives a synthetic PV which is in a very good agreement with the observational one, since in the locations that -according to the contour map- there is a knot, there is also a knot in the produced PV, too. This result clearly verifies the assumption for the "knots" model, this time from a spectrum-analysis perspective.

\begin{table*}
\centering
\caption{ Velocities and Kinematical ages of the torus and the different knots. The distance of each knot from the nebular center as long as the projected angle of ejection of each knot with respect to the main axis, as derived from the SHAPE model, are also presented. The systemic velocity is $V_{\rm sys}$ = -59~\,km\,s$^{-1}$.}
\label{tab:Table1}
\begin{tabular}{lrrrrrr} % five columns, alignment for each
\hline
Component  &Distance   &Angle &$V_{\rm hel}$    &$V_{\rm exp}$   &$V_{\rm out}$ & Kinematical age \\
           &(arcsec) &(degrees) & (\vel) &(\vel)   & (\vel) & (years)  \\ 
\hline
torus  & &           &$-72,-44$  &$-13,15$  &$15\pm2$     &$3190\pm1090$ \\
knot 1 &$14.87\pm1.48$ &4.8  &$ 2.4$  &$61$    &$185\pm15$     &$1100\pm390$ \\
knot 2 &$16.55\pm1.66$ &2.3  &$ 101$  &$160$   &$255\pm8$      &$887\pm305$ \\
knot 3 &$12.08\pm1.21$ &3.0  &$ 26.4$ &$85$    &$260\pm20$     &$635\pm224$ \\
knot 4 &$10.61\pm1.06$ &4.0  &$ 62$   &$121$   &$365\pm20$     &$397\pm140$ \\
knot 5 &$12.42\pm1.24$ &8.0  &$-160$  &$-101$  &$-390\pm10$    &$434\pm150$ \\
knot 6 &$13.88\pm1.40$ &8.3  &$-140$  &$-81$   &$-290\pm20$    &$660\pm232$ \\
knot 7 &$15.24\pm1.52$ &7.5  &$-117$  &$-58$   &$-185\pm25$    &$1125\pm417$ \\
knot 8 &$16.38\pm1.64$ &2.0  &$-219$  &$-160$  &$-251\pm10$    &$890\pm309$ \\
\hline
 \\ 

\end{tabular}
\end{table*}

\section{Discussion}
 \label{Discussion}

In this paper, we study the morpho-kinematical structure of Hb4 PN using new echelle spectroscopic data, high resolution {\it HST} images and SHAPE code. Hb4 displays a complex, multi-polar inner structure with a pair of seemingly collimated outflows. A bright fragmented ring-like structure is apparent together with an ellipsoidal component in the direction northwest-southeast (Fig.\ref{fig:HST4SLITSWfigure} (a)). 

In addition, a number of small scale features (knots) and filaments are also identified. We reckon that they exhibit a low-ionization spectrum similar to LISs (\citealt{GON2001}; \citealt{AKRGON2016}). Similar kind of fragmented equatorial rings and equatorial LISs have also been found in several other PNe, e.g. the Eskimo nebula (\citealt{GARDIA2012}) or the Necklace nebula (\citealt{COR2011}). The photo-ionization front at the later evolutionary stage of PNe has been proposed to be responsible for the fragmentation of shells (\citealt{GARSEG2006}) as well as an interaction between a jet/collimated outflow with the circumstellar envelope (\citealt{AKA2015}).
\par
Using the morpho-kinematical code SHAPE, we managed to reproduce the central ring-like structure and its PV diagram assuming a toroidal component whose major axis is at an angle of 107$\degr$ with respect to the North. The torus expands with an absolute mean velocity of $14\pm5$\,km\,s$^{-1}$, comparable with the velocities found in toroidal structures in PNe with [WR]-type nucleus (between 15-30 \vel; \citealt{AKRLOP2012}; \citealt{AKRBOU2015}; \citealt{GOM2018}) and in agreement within the errors to the velocity given by \citet{LOP1997}. If we take into account its angular size derived from our SHAPE model of $7$ arcsec and we adopt the distance of 2.88~$\pm0.86$ kpc, a kinematical age of 3190$\pm$1090 years for the torus is found (see Table \ref{tab:Table1}).

Regarding the bipolar outflows, decline radial outflow velocity as a function of the distance from the nucleus is found, as can be seen in Figure \ref{fig:plot3_errors}. This velocity law is uncommon in PNe given that most of jets or outflows show a linear increase of velocity with distance (e.g. \citealt{RIE2002}; \citealt{VAY2009}; \citealt{COR1999}; \citealt{GARSEG1999}). Nevertheless, a decline velocity behaviour has been reported in HH34 (\citealt{DEV1997}) and IRAS 18113-2503 (\citealt{ORO2018}) which is attributed to interaction of the outflows with the interstellar or circumstellar medium. This arose our curiosity to explore these bipolar outflows as a string of discrete knots rather than a jet. Making use of the {\it HST} \NII~6584 image and the echelle PV diagram, four knots in the northern outflow and three knots in the southern are identified. The resolution of our data may prevent the identification of more knots in the outflows, while the current picture of knotty outflows is different from the scenario of two elongated knots presented by \citet{LOP1997}. 

The knots are moving outward with a range of expansion velocities along the line of sight from  $-101$ \,km\,s$^{-1}$  to $160$~\,km\,s$^{-1}$, and a range of outflow velocities from $-390$ \,km\,s$^{-1}$ to $365$~\,km\,s$^{-1}$. The significant decline of outflow velocity with respect to the distance from the central star is also in conflict with the bow shock model presented by \citet{LOP1997}. Knot 2 is moving outward with an expansion velocity of $160$~ \,km\,s$^{-1}$, considerably higher than the remaining three knots in the northern and southern outflows and, according to the model, at an ejection angle from the main axis also higher than those of the rest knots, as illustrated in Figures \ref{fig:3da}(b), \ref{fig:4c}(c) and \ref{fig:5d}(d). This could be due to its ejection from the nebula during the rotation of the PN, as explained below. However, as it is shown in Figure \ref{fig:plot3_errors}, the outflow velocity of knot 2 is consistent with the radial decline outflow velocity law, which proves that it is travelling inside the same ISM as the rest knots do and experiences the same deceleration. For the ambient ISM, an isotropic gradient in its composition is assumed which accounts for the slowing down of the knots from the moment of their ejection.

The kinematical age of each knot in both outflows is estimated for D=2.88$\pm0.86$ kpc. From the results which are assembled in Table \ref{tab:Table1}, two significant conclusions can be drawn. The first is that, the knots in the pairs 1-7, 3-6, and 4-5 have the same kinematical age, which implies that the knot in each pair expelled at the same time. Knot 2, which as mentioned above, presents a displacement relative to the positions of the rest knots, had no equivalent component in the southern region included in slit S4 that could be considered its companion knot during the explosion from the core. So, we selected from the {\it HST} image a knot (knot 8). Since this knot was not included in slit S4, we could not have kinematical information about it. Thus, we assumed that knot 8 has the same $V_{\rm exp}$  and the same displacement parameter z as knot 2. Subsequently, the other two spacial parameters (x and y) of knot 8 were measured on the {\it HST} image and used in SHAPE. Using these assumptions and information we calculated its kinematical age. Indeed, its age is pretty much close to the age of knot 2, which allowed us to consider that knot 2 and knot 8 is a fourth pair of knots in Hb4. The second interesting comment regards a periodicity in the explosion events that the pairs of knots appear to have, since, according to the calculated kinematical ages, each pair of knots was expelled every $200-250$ years from the core of the PN. Given that the deceleration of the knots is present, the periodicity in the ejections could not have been perceptible if the composition of the ISM was irregularly inhomogeneous. Thus, the isotropic gradient in its composition mentioned above is supported which causes the deceleration of the knots, but also preserves the time interval between the motion of the knots, i.e. they undergo a similar deceleration. However, it should be taken into consideration that, the absolute values of the kinematical ages could be lower if there was no deceleration of the knots.
\par
In Figure \ref{fig:profiles_torus} the profiles of the seven knots found in slit S4 are presented along with the profile of the torus of the PN. It should be noted that there is a possibility that some of the knots might consist of more than one sub-knots which cannot be clearly identified due to the spectroscopic resolution. Representative example of nearby knots presented in the same profile, is that of knot 3 and knot 4 as shown in Figure \ref{fig:profiles_torus}(d). A second common feature in these profiles is the different background which is a clue for the inhomogeneity of the ISM that each knot encountered while travelling outwards the main core, resulting in the various expansion velocities of the knots as presented in Table \ref{tab:Table1}. In the torus profile, the two peaks correspond to the northern and southern part of the torus which is included in the slit S4, as this is shown in Figure \ref{fig:1c}(a). The heliocentric velocities of these parts are $-72$ \vel and  $-44$ \vel respectively, while their expansion velocities are $-13$ \vel and $15$ \vel respectively (see Table \ref{tab:Table1}). This difference in the velocities is justified due to the inclination of the torus relative to the line of sight. Thus, the absolute mean expansion velocity of the torus has been calculated at $14~$\,km\,s$^{-1}$  along the line of sight.
\par
The strong \NII~ emission detected in these knots compared to the value of the central parts implies some shock interaction with the circumstellar envelope (e.g. \citealt{RAG2008}; \citealt{AKRGON2016}; \citealt{AKRCLY2016}; \citealt{AKA2015}). By examining the high resolution {\it HST} \NII~image of Hb4, a misalignment of $\sim5$ degrees between the northern and southern string of knots is found. One possible interpretation for this, could be the ejection from a compact disk around a star in a binary that received a "push" by the wind from the other (p)AGB star \citep{HUA2013}, which eventually caused this misalignment in the emission of the knots. Another probable scenario could be a rotation mechanism present in its central stars. Very recently, \citet{GARSEG2014, GARSEG2016} have shown that surface rotational velocity could in principle produce considerable asymmetries in these stars. Hence, we suggest that each pair of knots was expelled from the core during the same outburst and, due to the above mentioned possible mechanisms, a few degrees of misalignment was created. Taking into account the periodicity in the emission of knots, the scenario of the rotating core is considered as most probable. This leads us to the suggestion that Hb4 is the case of a rotating central star which emits knots in an almost regular rate. Same mechanisms have been proposed to explain the knotty curved jets detected in NGC~6778 (\citealt{GUEMIR2012}) or Fleming~1 (\citealt{BOF2012}). \citet{LOP1993} have shown that the North and South strings of ionized knots in Fleming 1 PN could be generated by globules of gas ejected from a bipolar, rotating source within the core of Fleming 1. The BRETs scenario (\citealt{LOP1995}) provides plausible explanation for the formation of the  knots in Hb4.  All the above cases reinforce the hypothesis of a binary central system for this PN. However, the possibility that the knots had been shed symmetrically in the beginning, but the different ISM they encountered altered their velocity and inclination with respect to the initial emission axis, cannot be ruled out yet and further kinematicalalal observations are needed.

Besides the bipolar knots, the presence of the fragmented torus and the equatorial LISs may indicate of a PN formed via the common envelopes channel (\citealt{MIS2009}; \citealt{JON2014}; \citealt{GARSEG2018}). The fragmentation of the equatorial disk or torus into knots and filaments has been revealed from high resolution H$_2$ images of the bipolar PN NGC~2346 with a binary central system (\citealt{MAN2015}). \citet{AKRGON2017} have 
also confirmed the presence of H$_2$ gas in LISs embedded in NGC~7662 but there is no direct confirmation of a binary nucleus.

The classification of Hb4's central star as a hydrogen deficient [WR]-type (\citealt{ACKNEI2003}; \citealt{GOR2004}) makes it unique since none
of the known post-common envelope binary systems in PNe has a [WR] companion. The dual-dust chemistry reported for Hb4 (\citealt{PERCAL2009})
is also one of expected outcomes for a binary nucleus with [WR] component. Hb4 also belongs to the group of PNe with high ADF(O$^{+2}$) (3.7; \citealt{GARROJ2013}) which may also related with the presence of a binary central system. \citet{SOK1994} discuss the presence of
an accretion disk in PNe with binary central systems as one possible mechanism for the formation of highly collimated outflows/jets.
Jet interaction with circumstellar envelope can also lead to the formation of an equatorial torus as in Hb4 (\citealt{SOK2000}; \citealt{AKA2015}). \citet{AKA2008} confirmed that jets can be responsible of equatorial ring, but with a mass significantly less that the total
mass of the nebula, otherwise an additional mechanism is necessary for the formation of the torus e.g. common envelope interaction. An interesting
point is the finding that the jets appear to be younger than the torus in the hydrodynamic models despite they are formed by the same event (\citealt{AKA2008}). Indeed, the torus in Hb4 is found to be older than the knots (Table \ref{tab:Table1}).
In case of known post common-envelope PNe with binary nuclei, the toroidal component is younger than the jets (e.g. Ethos 1, \citealt{MIS2011}).
But, there are two more PNe (NGC~6337, NGC~6778; \citealt{TOCK2014}) in which the torus seems to have been formed before the jets. The delay between the equatorial torus and jets in Hb4 is of the order of 2000~years, comparable with the delay measured in NGC~6778.

The multi-polarity of Hb4 is also a very interesting finding as well as the various features which can be clearly seen in Figure \ref{fig:HST4SLITSWfigure}. \citet{RAB2003} reported a faint secondary bipolar structure, close to the central part, aligned with the minor axis.  Apart from that, Hb4 also displays two more bow-shaped filamentary structures in the northwest and southeast direction, formed probably from a different episodic event than those of the pair of knots. No kinematical information are available for these structures, so it is not possible to estimate the time they were formed. PNe with multi-polar inner structures and knotty outflows like Hb4 have also been found in other PNe (i.e. \citealt{PAL1996}; \citealt{HAR2004}). All these structures, their inclinations and morphologies implies that there have been a number of ejections during the lifetime of this PN.
However, further observations are needed (both high-resolution images and echelle spectra) in order to measure proper motions, velocities and the dynamical ages of these features and confirm how many different sequential ejections have been made providing valuable information 
about the formation and evolution of this PN with a possible binary nucleus.
 
 \section{Conclusion}
 
 In this paper a morpho-kinematical interpretation of the structure of the PN Hb4 is presented. We conclude the following:
 
 (1) The two bipolar outflows of Hb4 are classified as string of knots and characterized by the decline in their outflow velocity relative to their distance from the central star of the nebula. This deceleration is probably due to the interaction of the knots with an interstellar medium which presents an isotropic gradient in its composition, or nebular material in the vicinity of the nebula.
 
  (2) In the northern and southern outflows, four and three knots respectively were identified, each one of which travels outwards from the nebula with its own expansion velocity, covering a range from $-101~$\,km\,s$^{-1}$ up to $160~$\,km\,s$^{-1}$. The knots seem to have been expelled from the core in pairs, following a periodicity in the explosion events of $200-250$ years.
  
 (3) By the use of the code SHAPE, the central part of the PN was reconstructed as a torus, and found to have an absolute mean expansion velocity of $14~$\,km\,s$^{-1}$ along the line of sight. The low-ionization structures that the latter exhibits, indicate a common-envelope PN evolution.
 
 (4) The central part of the nebula is proposed to consist of a binary system having a [WR] companion evolved through the common envelopes channel.

 \section*{Acknowledgements}
  S.D. acknowledges the support of this work by (a) the PROTEAS II project (MIS 5002515), which is implemented under the "Reinforcement of the Research and Innovation Infrastructure" action, funded by the "Competitiveness, Entrepreneurship and Innovation" operational programme (NSRF 2014-2020) and co-financed by Greece and the European Union (European Regional Development Fund), and (b) the Operational Programme <<Human Resources Development, Education and Lifelong Learning>> in the context of the project ``Strengthening Human Resources Research Potential via Doctorate Research'' (MIS-5000432), implemented by the State Scholarships Foundation (IKY) and co-financed by Greece and the European Union (European Social Fund- ESF). S.A. acknowledges the financial support of the Brazilian agency Coordena\c{c}\~{a}o de Aperfei\c{c}oamento de Pessoal de N\'{i}vel Superior (CAPES) for a fellowship from the National Postdoctoral Program (PNPD) and Conselho Nacional de Desenvolvimento Cient\'{i}fico e Tecnol\'{o}gico 
(CNPq) through the grant 300336/2016-0). W.S. was supported by a grant from UNAM PAPIIT 104017.
This paper is based upon observations carried out at the Observatorio Astron\'{o}mico Nacional on the Sierra San Pedro M\'{a}rtir (OAN-SPM), Baja California, M\'{e}xico. Some of the data presented in this paper were obtained from the Mikulski Archive for Space Telescopes (MAST). 
STScI is operated by the Association of Universities for Research in Astronomy, Inc., under NASA contract NAS5-26555. 
 
%\subsection{Figures and tables}

%Figures and tables should be placed at logical positions in the text. Don't
%worry about the exact layout, which will be handled by the publishers.

%Figures are referred to as e.g. Fig.~\ref{fig:example_figure}, and tables as
%e.g. Table~\ref{tab:example_table}.

%\begin{table}
%	\centering
%	\caption{This is an example table. Captions appear above each table.
%	Remember to define the quantities, symbols and units used.}
%	\label{tab:example_table}
%	\begin{tabular}{lccr} % four columns, alignment for each
%		\hline
%		A & B & C & D\\
%		\hline
%		1 & 2 & 3 & 4\\
%		2 & 4 & 6 & 8\\
%		3 & 5 & 7 & 9\\
%		\hline
%	\end{tabular}
%\end{table}

%\section{Conclusions}

%The last numbered section should briefly summarise what has been done, and describe
%the final conclusions which the authors draw from their work.

%\section*{Acknowledgements}

%The Acknowledgements section is not numbered. Here you can thank helpful
%colleagues, acknowledge funding agencies, telescopes and facilities used etc.

%%%%%%%%%%%%%%%%%%%%%%%%%%%%%%%%%%%%%%%%%%%%%%%%%%


\bibliographystyle{mnras}
\bibliography{REF_HB4.bib} % if your bibtex file is called example.bib

\begin{thebibliography}{}
\makeatletter
\relax
\def\mn@urlcharsother{\let\do\@makeother \do\$\do\&\do\#\do\^\do\_\do\%\do\~}
\def\mn@doi{\begingroup\mn@urlcharsother \@ifnextchar [ {\mn@doi@}
  {\mn@doi@[]}}
\def\mn@doi@[#1]#2{\def\@tempa{#1}\ifx\@tempa\@empty \href
  {http://dx.doi.org/#2} {doi:#2}\else \href {http://dx.doi.org/#2} {#1}\fi
  \endgroup}
\def\mn@eprint#1#2{\mn@eprint@#1:#2::\@nil}
\def\mn@eprint@arXiv#1{\href {http://arxiv.org/abs/#1} {{\tt arXiv:#1}}}
\def\mn@eprint@dblp#1{\href {http://dblp.uni-trier.de/rec/bibtex/#1.xml}
  {dblp:#1}}
\def\mn@eprint@#1:#2:#3:#4\@nil{\def\@tempa {#1}\def\@tempb {#2}\def\@tempc
  {#3}\ifx \@tempc \@empty \let \@tempc \@tempb \let \@tempb \@tempa \fi \ifx
  \@tempb \@empty \def\@tempb {arXiv}\fi \@ifundefined
  {mn@eprint@\@tempb}{\@tempb:\@tempc}{\expandafter \expandafter \csname
  mn@eprint@\@tempb\endcsname \expandafter{\@tempc}}}

\bibitem[\protect\citeauthoryear{{Acker} \& {Neiner}}{{Acker} \&
  {Neiner}}{2003}]{ACKNEI2003}
{Acker} A.,  {Neiner} C.,  2003, \mn@doi [\aap] {10.1051/0004-6361:20030391},
  \href {http://adsabs.harvard.edu/abs/2003A%26A...403..659A} {403, 659}

\bibitem[\protect\citeauthoryear{{Akashi} \& {Soker}}{{Akashi} \&
  {Soker}}{2008}]{AKA2008}
{Akashi} M.,  {Soker} N.,  2008, \mn@doi [\mnras]
  {10.1111/j.1365-2966.2008.13935.x}, \href
  {http://adsabs.harvard.edu/abs/2008MNRAS.391.1063A} {391, 1063}

\bibitem[\protect\citeauthoryear{{Akashi}, {Sabach}, {Yogev}  \&
  {Soker}}{{Akashi} et~al.}{2015}]{AKA2015}
{Akashi} M.,  {Sabach} E.,  {Yogev} O.,   {Soker} N.,  2015, \mn@doi [\mnras]
  {10.1093/mnras/stv1666}, \href
  {http://adsabs.harvard.edu/abs/2015MNRAS.453.2115A} {453, 2115}

\bibitem[\protect\citeauthoryear{{Akras} \& {Gon{\c c}alves}}{{Akras} \&
  {Gon{\c c}alves}}{2016}]{AKRGON2016}
{Akras} S.,  {Gon{\c c}alves} D.~R.,  2016, \mn@doi [\mnras]
  {10.1093/mnras/stv2139}, \href
  {http://adsabs.harvard.edu/abs/2016MNRAS.455..930A} {455, 930}

\bibitem[\protect\citeauthoryear{{Akras} \& {L{\'o}pez}}{{Akras} \&
  {L{\'o}pez}}{2012}]{AKRLOP2012}
{Akras} S.,  {L{\'o}pez} J.~A.,  2012, \mn@doi [\mnras]
  {10.1111/j.1365-2966.2012.21578.x}, \href
  {http://adsabs.harvard.edu/abs/2012MNRAS.425.2197A} {425, 2197}

\bibitem[\protect\citeauthoryear{{Akras} \& {Steffen}}{{Akras} \&
  {Steffen}}{2012}]{AKRSTE2012}
{Akras} S.,  {Steffen} W.,  2012, \mn@doi [\mnras]
  {10.1111/j.1365-2966.2012.20928.x}, \href
  {http://adsabs.harvard.edu/abs/2012MNRAS.423..925A} {423, 925}

\bibitem[\protect\citeauthoryear{{Akras}, {Boumis}, {Meaburn}, {Alikakos},
  {L{\'o}pez}  \& {Gon{\c c}alves}}{{Akras} et~al.}{2015}]{AKRBOU2015}
{Akras} S.,  {Boumis} P.,  {Meaburn} J.,  {Alikakos} J.,  {L{\'o}pez} J.~A.,
  {Gon{\c c}alves} D.~R.,  2015, \mn@doi [\mnras] {10.1093/mnras/stv1468},
  \href {http://adsabs.harvard.edu/abs/2015MNRAS.452.2911A} {452, 2911}

\bibitem[\protect\citeauthoryear{{Akras}, {Clyne}, {Boumis}, {Monteiro},
  {Gon{\c c}alves}, {Redman}  \& {Williams}}{{Akras} et~al.}{2016}]{AKRCLY2016}
{Akras} S.,  {Clyne} N.,  {Boumis} P.,  {Monteiro} H.,  {Gon{\c c}alves} D.~R.,
   {Redman} M.~P.,   {Williams} S.,  2016, \mn@doi [\mnras]
  {10.1093/mnras/stw038}, \href
  {http://adsabs.harvard.edu/abs/2016MNRAS.457.3409A} {457, 3409}

\bibitem[\protect\citeauthoryear{{Akras}, {Gon{\c c}alves}  \&
  {Ramos-Larios}}{{Akras} et~al.}{2017}]{AKRGON2017}
{Akras} S.,  {Gon{\c c}alves} D.~R.,   {Ramos-Larios} G.,  2017, \mn@doi
  [\mnras] {10.1093/mnras/stw2736}, \href
  {http://adsabs.harvard.edu/abs/2017MNRAS.465.1289A} {465, 1289}

\bibitem[\protect\citeauthoryear{{Bailer-Jones}, {Rybizki}, {Fouesneau},
  {Mantelet}  \& {Andrae}}{{Bailer-Jones} et~al.}{2018}]{BAILJON2018}
{Bailer-Jones} C.~A.~L.,  {Rybizki} J.,  {Fouesneau} M.,  {Mantelet} G.,
  {Andrae} R.,  2018, \mn@doi [\aj] {10.3847/1538-3881/aacb21}, \href
  {http://cdsads.u-strasbg.fr/abs/2018AJ....156...58B} {156, 58}

\bibitem[\protect\citeauthoryear{{Balick}}{{Balick}}{1987}]{BAL1987}
{Balick} B.,  1987, \mn@doi [\aj] {10.1086/114504}, \href
  {http://adsabs.harvard.edu/abs/1987AJ.....94..671B} {94, 671}

\bibitem[\protect\citeauthoryear{{Balick}, {Rugers}, {Terzian}  \&
  {Chengalur}}{{Balick} et~al.}{1993}]{BAL1993}
{Balick} B.,  {Rugers} M.,  {Terzian} Y.,   {Chengalur} J.~N.,  1993, \mn@doi
  [\apj] {10.1086/172881}, \href
  {http://adsabs.harvard.edu/abs/1993ApJ...411..778B} {411, 778}

\bibitem[\protect\citeauthoryear{{Balick}, {Alexander}, {Hajian}, {Terzian},
  {Perinotto}  \& {Patriarchi}}{{Balick} et~al.}{1998}]{BAL1998}
{Balick} B.,  {Alexander} J.,  {Hajian} A.~R.,  {Terzian} Y.,  {Perinotto} M.,
   {Patriarchi} P.,  1998, \mn@doi [\aj] {10.1086/300429}, \href
  {http://adsabs.harvard.edu/abs/1998AJ....116..360B} {116, 360}

\bibitem[\protect\citeauthoryear{{Boffin}, {Miszalski}, {Rauch}, {Jones},
  {Corradi}, {Napiwotzki}, {Day-Jones}  \& {K{\"o}ppen}}{{Boffin}
  et~al.}{2012}]{BOF2012}
{Boffin} H.~M.~J.,  {Miszalski} B.,  {Rauch} T.,  {Jones} D.,  {Corradi}
  R.~L.~M.,  {Napiwotzki} R.,  {Day-Jones} A.~C.,   {K{\"o}ppen} J.,  2012,
  \mn@doi [Science] {10.1126/science.1225386}, \href
  {http://adsabs.harvard.edu/abs/2012Sci...338..773B} {338, 773}

\bibitem[\protect\citeauthoryear{{Borkowski}}{{Borkowski}}{1996}]{BOR1996}
{Borkowski} K.,  1996, {A Search for Jets in Planetary Nebulae}, HST Proposal

\bibitem[\protect\citeauthoryear{{Boumis}, {Paleologou}, {Mavromatakis}  \&
  {Papamastorakis}}{{Boumis} et~al.}{2003}]{BOUMIS2003}
{Boumis} P.,  {Paleologou} E.~V.,  {Mavromatakis} F.,   {Papamastorakis} J.,
  2003, \mn@doi [\mnras] {10.1046/j.1365-8711.2003.06233.x}, \href
  {http://adsabs.harvard.edu/abs/2003MNRAS.339..735B} {339, 735}

\bibitem[\protect\citeauthoryear{{Boumis}, {Akras}, {Xilouris}, {Mavromatakis},
  {Kapakos}, {Papamastorakis}  \& {Goudis}}{{Boumis} et~al.}{2006}]{BOUMIS2006}
{Boumis} P.,  {Akras} S.,  {Xilouris} E.~M.,  {Mavromatakis} F.,  {Kapakos} E.,
   {Papamastorakis} J.,   {Goudis} C.~D.,  2006, \mn@doi [\mnras]
  {10.1111/j.1365-2966.2006.10048.x}, \href
  {http://adsabs.harvard.edu/abs/2006MNRAS.367.1551B} {367, 1551}

\bibitem[\protect\citeauthoryear{{Clyne}, {Akras}, {Steffen}, {Redman}, {Gon{\c
  c}alves}  \& {Harvey}}{{Clyne} et~al.}{2015}]{CLY2015}
{Clyne} N.,  {Akras} S.,  {Steffen} W.,  {Redman} M.~P.,  {Gon{\c c}alves}
  D.~R.,   {Harvey} E.,  2015, \mn@doi [\aap] {10.1051/0004-6361/201526585},
  \href {http://adsabs.harvard.edu/abs/2015A%26A...582A..60C} {582, A60}

\bibitem[\protect\citeauthoryear{{Corradi}, {Manso}, {Mampaso}  \&
  {Schwarz}}{{Corradi} et~al.}{1996}]{COR1996}
{Corradi} R.~L.~M.,  {Manso} R.,  {Mampaso} A.,   {Schwarz} H.~E.,  1996, \aap,
  \href {http://adsabs.harvard.edu/abs/1996A%26A...313..913C} {313, 913}

\bibitem[\protect\citeauthoryear{{Corradi}, {Perinotto}, {Villaver}, {Mampaso}
  \& {Gon{\c c}alves}}{{Corradi} et~al.}{1999}]{COR1999}
{Corradi} R.~L.~M.,  {Perinotto} M.,  {Villaver} E.,  {Mampaso} A.,   {Gon{\c
  c}alves} D.~R.,  1999, \mn@doi [\apj] {10.1086/307768}, \href
  {http://adsabs.harvard.edu/abs/1999ApJ...523..721C} {523, 721}

\bibitem[\protect\citeauthoryear{{Corradi} et~al.,}{{Corradi}
  et~al.}{2011}]{COR2011}
{Corradi} R.~L.~M.,  et~al., 2011, \mn@doi [\mnras]
  {10.1111/j.1365-2966.2010.17523.x}, \href
  {http://adsabs.harvard.edu/abs/2011MNRAS.410.1349C} {410, 1349}

\bibitem[\protect\citeauthoryear{{Corradi}, {Garc{\'{\i}}a-Rojas}, {Jones}  \&
  {Rodr{\'{\i}}guez-Gil}}{{Corradi} et~al.}{2015}]{COR2015}
{Corradi} R.~L.~M.,  {Garc{\'{\i}}a-Rojas} J.,  {Jones} D.,
  {Rodr{\'{\i}}guez-Gil} P.,  2015, \mn@doi [\apj]
  {10.1088/0004-637X/803/2/99}, \href
  {http://adsabs.harvard.edu/abs/2015ApJ...803...99C} {803, 99}

\bibitem[\protect\citeauthoryear{{Danehkar}}{{Danehkar}}{2014}]{DAN2014}
{Danehkar} A.,  2014, PhD thesis, Macquarie University, Australia,
  \mn@doi{10.5281/zenodo.47794}

\bibitem[\protect\citeauthoryear{{Devine}, {Bally}, {Reipurth}  \&
  {Heathcote}}{{Devine} et~al.}{1997}]{DEV1997}
{Devine} D.,  {Bally} J.,  {Reipurth} B.,   {Heathcote} S.,  1997, \mn@doi
  [\aj] {10.1086/118629}, \href
  {http://adsabs.harvard.edu/abs/1997AJ....114.2095D} {114, 2095}

\bibitem[\protect\citeauthoryear{{Frank}, {Chen}, {Reichardt}, {De Marco},
  {Blackman}  \& {Nordhaus}}{{Frank} et~al.}{2018}]{FRA2018}
{Frank} A.,  {Chen} Z.,  {Reichardt} T.,  {De Marco} O.,  {Blackman} E.,
  {Nordhaus} J.,  2018, preprint, \href
  {http://adsabs.harvard.edu/abs/2018arXiv180705925F} {} (\mn@eprint {arXiv}
  {1807.05925})

\bibitem[\protect\citeauthoryear{{Frew}, {Parker}  \& {Boji{\v
  c}i{\'c}}}{{Frew} et~al.}{2016}]{FRE2016}
{Frew} D.~J.,  {Parker} Q.~A.,   {Boji{\v c}i{\'c}} I.~S.,  2016, \mn@doi
  [\mnras] {10.1093/mnras/stv1516}, \href
  {http://adsabs.harvard.edu/abs/2016MNRAS.455.1459F} {455, 1459}

\bibitem[\protect\citeauthoryear{{Gaia Collaboration}}{{Gaia
  Collaboration}}{2018}]{GAI2018}
{Gaia Collaboration} 2018, VizieR Online Data Catalog, \href
  {http://adsabs.harvard.edu/abs/2018yCat.1345....0G} {1345}

\bibitem[\protect\citeauthoryear{{Garc{\'{\i}}a-D{\'{\i}}az}, {L{\'o}pez},
  {Steffen}  \& {Richer}}{{Garc{\'{\i}}a-D{\'{\i}}az}
  et~al.}{2012}]{GARDIA2012}
{Garc{\'{\i}}a-D{\'{\i}}az} M.~T.,  {L{\'o}pez} J.~A.,  {Steffen} W.,
  {Richer} M.~G.,  2012, \mn@doi [\apj] {10.1088/0004-637X/761/2/172}, \href
  {http://adsabs.harvard.edu/abs/2012ApJ...761..172G} {761, 172}

\bibitem[\protect\citeauthoryear{{Garc{\'{\i}}a-Rojas}, {Pe{\~n}a}, {Morisset},
  {Delgado-Inglada}, {Mesa-Delgado}  \& {Ruiz}}{{Garc{\'{\i}}a-Rojas}
  et~al.}{2013}]{GARROJ2013}
{Garc{\'{\i}}a-Rojas} J.,  {Pe{\~n}a} M.,  {Morisset} C.,  {Delgado-Inglada}
  G.,  {Mesa-Delgado} A.,   {Ruiz} M.~T.,  2013, \mn@doi [\aap]
  {10.1051/0004-6361/201322354}, \href
  {http://adsabs.harvard.edu/abs/2013A%26A...558A.122G} {558, A122}

\bibitem[\protect\citeauthoryear{{Garc{\'{\i}}a-Segura}, {Langer},
  {R{\'o}{\.z}yczka}  \& {Franco}}{{Garc{\'{\i}}a-Segura}
  et~al.}{1999}]{GARSEG1999}
{Garc{\'{\i}}a-Segura} G.,  {Langer} N.,  {R{\'o}{\.z}yczka} M.,   {Franco} J.,
   1999, \mn@doi [\apj] {10.1086/307205}, \href
  {http://adsabs.harvard.edu/abs/1999ApJ...517..767G} {517, 767}

\bibitem[\protect\citeauthoryear{{Garc{\'{\i}}a-Segura}, {L{\'o}pez},
  {Steffen}, {Meaburn}  \& {Manchado}}{{Garc{\'{\i}}a-Segura}
  et~al.}{2006}]{GARSEG2006}
{Garc{\'{\i}}a-Segura} G.,  {L{\'o}pez} J.~A.,  {Steffen} W.,  {Meaburn} J.,
  {Manchado} A.,  2006, \mn@doi [\apjl] {10.1086/506559}, \href
  {http://adsabs.harvard.edu/abs/2006ApJ...646L..61G} {646, L61}

\bibitem[\protect\citeauthoryear{{Garc{\'{\i}}a-Segura}, {Villaver}, {Langer},
  {Yoon}  \& {Manchado}}{{Garc{\'{\i}}a-Segura} et~al.}{2014}]{GARSEG2014}
{Garc{\'{\i}}a-Segura} G.,  {Villaver} E.,  {Langer} N.,  {Yoon} S.-C.,
  {Manchado} A.,  2014, \mn@doi [\apj] {10.1088/0004-637X/783/2/74}, \href
  {http://adsabs.harvard.edu/abs/2014ApJ...783...74G} {783, 74}

\bibitem[\protect\citeauthoryear{{Garc{\'{\i}}a-Segura}, {Villaver},
  {Manchado}, {Langer}  \& {Yoon}}{{Garc{\'{\i}}a-Segura}
  et~al.}{2016}]{GARSEG2016}
{Garc{\'{\i}}a-Segura} G.,  {Villaver} E.,  {Manchado} A.,  {Langer} N.,
  {Yoon} S.-C.,  2016, \mn@doi [\apj] {10.3847/0004-637X/823/2/142}, \href
  {http://adsabs.harvard.edu/abs/2016ApJ...823..142G} {823, 142}

\bibitem[\protect\citeauthoryear{{Garc{\'{\i}}a-Segura}, {Ricker}  \&
  {Taam}}{{Garc{\'{\i}}a-Segura} et~al.}{2018}]{GARSEG2018}
{Garc{\'{\i}}a-Segura} G.,  {Ricker} P.~M.,   {Taam} R.~E.,  2018, \mn@doi
  [\apj] {10.3847/1538-4357/aac08c}, \href
  {http://adsabs.harvard.edu/abs/2018ApJ...860...19G} {860, 19}

\bibitem[\protect\citeauthoryear{{G{\'o}mez}, {Niccolini}, {Su{\'a}rez},
  {Miranda}, {Rizzo}, {Uscanga}, {Green}  \& {de
  Gregorio-Monsalvo}}{{G{\'o}mez} et~al.}{2018}]{GOM2018}
{G{\'o}mez} J.~F.,  {Niccolini} G.,  {Su{\'a}rez} O.,  {Miranda} L.~F.,
  {Rizzo} J.~R.,  {Uscanga} L.,  {Green} J.~A.,   {de Gregorio-Monsalvo} I.,
  2018, \mn@doi [\mnras] {10.1093/mnras/sty2193}, \href
  {http://adsabs.harvard.edu/abs/2018MNRAS.480.4991G} {480, 4991}

\bibitem[\protect\citeauthoryear{{Gon{\c c}alves}, {Corradi}  \&
  {Mampaso}}{{Gon{\c c}alves} et~al.}{2001}]{GON2001}
{Gon{\c c}alves} D.~R.,  {Corradi} R.~L.~M.,   {Mampaso} A.,  2001, \mn@doi
  [\apj] {10.1086/318364}, \href
  {http://adsabs.harvard.edu/abs/2001ApJ...547..302G} {547, 302}

\bibitem[\protect\citeauthoryear{{G{\'o}rny}, {Stasi{\'n}ska}, {Escudero}  \&
  {Costa}}{{G{\'o}rny} et~al.}{2004}]{GOR2004}
{G{\'o}rny} S.~K.,  {Stasi{\'n}ska} G.,  {Escudero} A.~V.,   {Costa} R.~D.~D.,
  2004, \mn@doi [\aap] {10.1051/0004-6361:20047064}, \href
  {http://adsabs.harvard.edu/abs/2004A%26A...427..231G} {427, 231}

\bibitem[\protect\citeauthoryear{{Guerrero} \& {Miranda}}{{Guerrero} \&
  {Miranda}}{2012}]{GUEMIR2012}
{Guerrero} M.~A.,  {Miranda} L.~F.,  2012, \mn@doi [\aap]
  {10.1051/0004-6361/201117923}, \href
  {http://adsabs.harvard.edu/abs/2012A%26A...539A..47G} {539, A47}

\bibitem[\protect\citeauthoryear{{Hajian}, {Balick}, {Terzian}  \&
  {Perinotto}}{{Hajian} et~al.}{1997}]{HAJ1997}
{Hajian} A.~R.,  {Balick} B.,  {Terzian} Y.,   {Perinotto} M.,  1997, \mn@doi
  [\apj] {10.1086/304598}, \href
  {http://adsabs.harvard.edu/abs/1997ApJ...487..304H} {487, 304}

\bibitem[\protect\citeauthoryear{{Harman}, {Bryce}, {L{\'o}pez}, {Meaburn}  \&
  {Holloway}}{{Harman} et~al.}{2004}]{HAR2004}
{Harman} D.~J.,  {Bryce} M.,  {L{\'o}pez} J.~A.,  {Meaburn} J.,   {Holloway}
  A.~J.,  2004, \mn@doi [\mnras] {10.1111/j.1365-2966.2004.07427.x}, \href
  {http://adsabs.harvard.edu/abs/2004MNRAS.348.1047H} {348, 1047}

\bibitem[\protect\citeauthoryear{{Huarte-Espinosa}, {Carroll-Nellenback},
  {Nordhaus}, {Frank}  \& {Blackman}}{{Huarte-Espinosa} et~al.}{2013}]{HUA2013}
{Huarte-Espinosa} M.,  {Carroll-Nellenback} J.,  {Nordhaus} J.,  {Frank} A.,
  {Blackman} E.~G.,  2013, \mn@doi [\mnras] {10.1093/mnras/stt725}, \href
  {http://adsabs.harvard.edu/abs/2013MNRAS.433..295H} {433, 295}

\bibitem[\protect\citeauthoryear{{Icke}}{{Icke}}{1988}]{ICK1988}
{Icke} V.,  1988, \aap, \href
  {http://adsabs.harvard.edu/abs/1988A%26A...202..177I} {202, 177}

\bibitem[\protect\citeauthoryear{{Jones}, {Boffin}, {Miszalski}, {Wesson},
  {Corradi}  \& {Tyndall}}{{Jones} et~al.}{2014}]{JON2014}
{Jones} D.,  {Boffin} H.~M.~J.,  {Miszalski} B.,  {Wesson} R.,  {Corradi}
  R.~L.~M.,   {Tyndall} A.~A.,  2014, \mn@doi [\aap]
  {10.1051/0004-6361/201322797}, \href
  {http://adsabs.harvard.edu/abs/2014A%26A...562A..89J} {562, A89}

\bibitem[\protect\citeauthoryear{{Kwok}, {Purton}  \& {Fitzgerald}}{{Kwok}
  et~al.}{1978}]{KWO1978}
{Kwok} S.,  {Purton} C.~R.,   {Fitzgerald} P.~M.,  1978, \mn@doi [\apjl]
  {10.1086/182621}, \href {http://adsabs.harvard.edu/abs/1978ApJ...219L.125K}
  {219, L125}

\bibitem[\protect\citeauthoryear{{Livio} \& {Soker}}{{Livio} \&
  {Soker}}{1988}]{LIV1988}
{Livio} M.,  {Soker} N.,  1988, \mn@doi [\apj] {10.1086/166419}, \href
  {http://adsabs.harvard.edu/abs/1988ApJ...329..764L} {329, 764}

\bibitem[\protect\citeauthoryear{{Lopez}, {Meaburn}  \& {Palmer}}{{Lopez}
  et~al.}{1993}]{LOP1993}
{Lopez} J.~A.,  {Meaburn} J.,   {Palmer} J.~W.,  1993, \mn@doi [\apjl]
  {10.1086/187051}, \href {http://adsabs.harvard.edu/abs/1993ApJ...415L.135L}
  {415, L135}

\bibitem[\protect\citeauthoryear{{Lopez}, {Vazquez}  \& {Rodriguez}}{{Lopez}
  et~al.}{1995}]{LOP1995}
{Lopez} J.~A.,  {Vazquez} R.,   {Rodriguez} L.~F.,  1995, \mn@doi [\apjl]
  {10.1086/309801}, \href {http://adsabs.harvard.edu/abs/1995ApJ...455L..63L}
  {455, L63}

\bibitem[\protect\citeauthoryear{{L{\'o}pez}, {Steffen}  \&
  {Meaburn}}{{L{\'o}pez} et~al.}{1997}]{LOP1997}
{L{\'o}pez} J.~A.,  {Steffen} W.,   {Meaburn} J.,  1997, \mn@doi [\apj]
  {10.1086/304472}, \href {http://adsabs.harvard.edu/abs/1997ApJ...485..697L}
  {485, 697}

\bibitem[\protect\citeauthoryear{{Manchado}, {Stanghellini}, {Villaver},
  {Garc{\'{\i}}a-Segura}, {Shaw}  \& {Garc{\'{\i}}a-Hern{\'a}ndez}}{{Manchado}
  et~al.}{2015}]{MAN2015}
{Manchado} A.,  {Stanghellini} L.,  {Villaver} E.,  {Garc{\'{\i}}a-Segura} G.,
  {Shaw} R.~A.,   {Garc{\'{\i}}a-Hern{\'a}ndez} D.~A.,  2015, \mn@doi [\apj]
  {10.1088/0004-637X/808/2/115}, \href
  {http://adsabs.harvard.edu/abs/2015ApJ...808..115M} {808, 115}

\bibitem[\protect\citeauthoryear{{Meaburn}, {L{\'o}pez}, {Guti{\'e}rrez},
  {Quir{\'o}z}, {Murillo}, {Vald{\'e}z}  \& {Pedrayez}}{{Meaburn}
  et~al.}{2003}]{Meaburn2003}
{Meaburn} J.,  {L{\'o}pez} J.~A.,  {Guti{\'e}rrez} L.,  {Quir{\'o}z} F.,
  {Murillo} J.~M.,  {Vald{\'e}z} J.,   {Pedrayez} M.,  2003, \rmxaa, \href
  {http://adsabs.harvard.edu/abs/2003RMxAA..39..185M} {39, 185}

\bibitem[\protect\citeauthoryear{{Miszalski}, {Acker}, {Parker}  \&
  {Moffat}}{{Miszalski} et~al.}{2009}]{MIS2009}
{Miszalski} B.,  {Acker} A.,  {Parker} Q.~A.,   {Moffat} A.~F.~J.,  2009,
  \mn@doi [\aap] {10.1051/0004-6361/200912176}, \href
  {http://adsabs.harvard.edu/abs/2009A%26A...505..249M} {505, 249}

\bibitem[\protect\citeauthoryear{{Miszalski}, {Corradi}, {Boffin}, {Jones},
  {Sabin}, {Santander-Garc{\'{\i}}a}, {Rodr{\'{\i}}guez-Gil}  \&
  {Rubio-D{\'{\i}}ez}}{{Miszalski} et~al.}{2011}]{MIS2011}
{Miszalski} B.,  {Corradi} R.~L.~M.,  {Boffin} H.~M.~J.,  {Jones} D.,  {Sabin}
  L.,  {Santander-Garc{\'{\i}}a} M.,  {Rodr{\'{\i}}guez-Gil} P.,
  {Rubio-D{\'{\i}}ez} M.~M.,  2011, \mn@doi [\mnras]
  {10.1111/j.1365-2966.2011.18212.x}, \href
  {http://adsabs.harvard.edu/abs/2011MNRAS.413.1264M} {413, 1264}

\bibitem[\protect\citeauthoryear{{Orosz} et~al.,}{{Orosz}
  et~al.}{2018}]{ORO2018}
{Orosz} G.,  et~al., 2018, preprint, \href
  {http://adsabs.harvard.edu/abs/2018arXiv180907505O} {} (\mn@eprint {arXiv}
  {1809.07505})

\bibitem[\protect\citeauthoryear{{Palmer}, {Lopez}, {Meaburn}  \&
  {Lloyd}}{{Palmer} et~al.}{1996}]{PAL1996}
{Palmer} J.~W.,  {Lopez} J.~A.,  {Meaburn} J.,   {Lloyd} H.~M.,  1996, \aap,
  \href {http://adsabs.harvard.edu/abs/1996A%26A...307..225P} {307, 225}

\bibitem[\protect\citeauthoryear{{Pe{\~n}a}, {Ruiz-Escobedo},
  {Rechy-Garc{\'{\i}}a}  \& {Garc{\'{\i}}a-Rojas}}{{Pe{\~n}a}
  et~al.}{2017}]{PEN2017}
{Pe{\~n}a} M.,  {Ruiz-Escobedo} F.,  {Rechy-Garc{\'{\i}}a} J.~S.,
  {Garc{\'{\i}}a-Rojas} J.,  2017, \mn@doi [\mnras] {10.1093/mnras/stx1991},
  \href {http://adsabs.harvard.edu/abs/2017MNRAS.472.1182P} {472, 1182}

\bibitem[\protect\citeauthoryear{{Perea-Calder{\'o}n},
  {Garc{\'{\i}}a-Hern{\'a}ndez}, {Garc{\'{\i}}a-Lario}, {Szczerba}  \&
  {Bobrowsky}}{{Perea-Calder{\'o}n} et~al.}{2009}]{PERCAL2009}
{Perea-Calder{\'o}n} J.~V.,  {Garc{\'{\i}}a-Hern{\'a}ndez} D.~A.,
  {Garc{\'{\i}}a-Lario} P.,  {Szczerba} R.,   {Bobrowsky} M.,  2009, \mn@doi
  [\aap] {10.1051/0004-6361:200811457}, \href
  {http://adsabs.harvard.edu/abs/2009A%26A...495L...5P} {495, L5}

\bibitem[\protect\citeauthoryear{{Perinotto}}{{Perinotto}}{2000}]{PER2000}
{Perinotto} M.,  2000, \mn@doi [\apss] {10.1023/A:1026516527132}, \href
  {http://adsabs.harvard.edu/abs/2000Ap%26SS.274..205P} {274, 205}

\bibitem[\protect\citeauthoryear{{Raba{\c c}a}, {Cuisinier}, {Lorenz-Martins},
  {Epit{\'a}cio Pereira}, {Gon{\c c}alves}  \& {Lastennet}}{{Raba{\c c}a}
  et~al.}{2003}]{RAB2003}
{Raba{\c c}a} C.~R.,  {Cuisinier} F.,  {Lorenz-Martins} S.,  {Epit{\'a}cio
  Pereira} D.~N.,  {Gon{\c c}alves} D.~R.,   {Lastennet} E.,  2003, in {Kwok}
  S.,  {Dopita} M.,   {Sutherland} R.,  eds,  IAU Symposium Vol. 209, Planetary
  Nebulae: Their Evolution and Role in the Universe. p.~491

\bibitem[\protect\citeauthoryear{{Raga}, {Riera}, {Mellema}, {Esquivel}  \&
  {Vel{\'a}zquez}}{{Raga} et~al.}{2008}]{RAG2008}
{Raga} A.~C.,  {Riera} A.,  {Mellema} G.,  {Esquivel} A.,   {Vel{\'a}zquez}
  P.~F.,  2008, \mn@doi [\aap] {10.1051/0004-6361:20079157}, \href
  {http://adsabs.harvard.edu/abs/2008A%26A...489.1141R} {489, 1141}

\bibitem[\protect\citeauthoryear{{Riera}, {Garc{\'{\i}}a-Lario}, {Manchado},
  {Bobrowsky}  \& {Estalella}}{{Riera} et~al.}{2002}]{RIE2002}
{Riera} A.,  {Garc{\'{\i}}a-Lario} P.,  {Manchado} A.,  {Bobrowsky} M.,
  {Estalella} R.,  2002, in {Henney} W.~J.,  {Steffen} W.,  {Binette} L.,
  {Raga} A.,  eds,  Revista Mexicana de Astronomia y Astrofisica Conference
  Series Vol. 13, Revista Mexicana de Astronomia y Astrofisica Conference
  Series. pp 127--132

\bibitem[\protect\citeauthoryear{{Robinson}, {Reay}  \& {Atherton}}{{Robinson}
  et~al.}{1982}]{ROB1982}
{Robinson} G.~J.,  {Reay} N.~K.,   {Atherton} P.~D.,  1982, \mn@doi [\mnras]
  {10.1093/mnras/199.3.649}, \href
  {http://adsabs.harvard.edu/abs/1982MNRAS.199..649R} {199, 649}

\bibitem[\protect\citeauthoryear{{Sahai}, {Morris}  \& {Villar}}{{Sahai}
  et~al.}{2011}]{SAH2011}
{Sahai} R.,  {Morris} M.~R.,   {Villar} G.~G.,  2011, \mn@doi [\aj]
  {10.1088/0004-6256/141/4/134}, \href
  {http://adsabs.harvard.edu/abs/2011AJ....141..134S} {141, 134}

\bibitem[\protect\citeauthoryear{{Schwarz}, {Corradi}  \& {Melnick}}{{Schwarz}
  et~al.}{1992}]{SCH1992}
{Schwarz} H.~E.,  {Corradi} R.~L.~M.,   {Melnick} J.,  1992, \aaps, \href
  {http://adsabs.harvard.edu/abs/1992A%26AS...96...23S} {96, 23}

\bibitem[\protect\citeauthoryear{{Soker} \& {Livio}}{{Soker} \&
  {Livio}}{1994}]{SOK1994}
{Soker} N.,  {Livio} M.,  1994, \mn@doi [\apj] {10.1086/173639}, \href
  {http://adsabs.harvard.edu/abs/1994ApJ...421..219S} {421, 219}

\bibitem[\protect\citeauthoryear{{Soker} \& {Rappaport}}{{Soker} \&
  {Rappaport}}{2000}]{SOK2000}
{Soker} N.,  {Rappaport} S.,  2000, \mn@doi [\apj] {10.1086/309112}, \href
  {http://adsabs.harvard.edu/abs/2000ApJ...538..241S} {538, 241}

\bibitem[\protect\citeauthoryear{{Steffen} \& {L{\'o}pez}}{{Steffen} \&
  {L{\'o}pez}}{2006}]{STE2006}
{Steffen} W.,  {L{\'o}pez} J.~A.,  2006, \rmxaa, \href
  {http://adsabs.harvard.edu/abs/2006RMxAA..42...99S} {42, 99}

\bibitem[\protect\citeauthoryear{{Steffen}, {Koning}, {Wenger}, {Morisset}  \&
  {Magnor}}{{Steffen} et~al.}{2011}]{STE2011}
{Steffen} W.,  {Koning} N.,  {Wenger} S.,  {Morisset} C.,   {Magnor} M.,  2011,
  \mn@doi [IEEE Transactions on Visualization and Computer Graphics, Volume 17,
  Issue 4, p.454-465] {10.1109/TVCG.2010.62}, \href
  {http://adsabs.harvard.edu/abs/2011ITVCG..17..454S} {17, 454}

\bibitem[\protect\citeauthoryear{{Tocknell}, {De Marco}  \&
  {Wardle}}{{Tocknell} et~al.}{2014}]{TOCK2014}
{Tocknell} J.,  {De Marco} O.,   {Wardle} M.,  2014, \mn@doi [\mnras]
  {10.1093/mnras/stu079}, \href
  {http://adsabs.harvard.edu/abs/2014MNRAS.439.2014T} {439, 2014}

\bibitem[\protect\citeauthoryear{{Vaytet}, {Rushton}, {Lloyd}, {L{\'o}pez},
  {Meaburn}, {O'Brien}, {Mitchell}  \& {Pollacco}}{{Vaytet}
  et~al.}{2009}]{VAY2009}
{Vaytet} N.~M.~H.,  {Rushton} A.~P.,  {Lloyd} M.,  {L{\'o}pez} J.~A.,
  {Meaburn} J.,  {O'Brien} T.~J.,  {Mitchell} D.~L.,   {Pollacco} D.,  2009,
  \mn@doi [\mnras] {10.1111/j.1365-2966.2009.15149.x}, \href
  {http://adsabs.harvard.edu/abs/2009MNRAS.398..385V} {398, 385}

\bibitem[\protect\citeauthoryear{{Weidmann}, {Schmidt}, {Vena Valdarenas},
  {Ahumada}, {Volpe}  \& {Mudrik}}{{Weidmann} et~al.}{2016}]{WEI2016}
{Weidmann} W.~A.,  {Schmidt} E.~O.,  {Vena Valdarenas} R.~R.,  {Ahumada} J.~A.,
   {Volpe} M.~G.,   {Mudrik} A.,  2016, \mn@doi [\aap]
  {10.1051/0004-6361/201527199}, \href
  {http://adsabs.harvard.edu/abs/2016A%26A...592A.103W} {592, A103}

\makeatother
\end{thebibliography}

%\begin{thebibliography}{99}
%\end{thebibliography}

%%%%%%%%%%%%%%%%%%%%%%%%%%%%%%%%%%%%%%%%%%%%%%%%%%

%%%%%%%%%%%%%%%%% APPENDICES %%%%%%%%%%%%%%%%%%%%%

%%%%%%%%%%%%%%%%%%%%%%%%%%%%%%%%%%%%%%%%%%%%%%%%%%

% Don't change these lines
\bsp	% typesetting comment
\label{lastpage}
\end{document}